\documentclass[a4paper,11pt]{article}

\usepackage{geometry}
\usepackage{amsmath}
\usepackage{graphicx}
\usepackage{braket}
\usepackage{float}
\usepackage{caption}
\usepackage{subcaption}
\usepackage{url}
\usepackage{csquotes}
\usepackage{longtable}
\usepackage{amssymb}
\usepackage{bbm}

\usepackage{xcolor}
\usepackage[backend=biber,style=numeric-comp,sorting=none]{biblatex}

\usepackage{tikz}
\usetikzlibrary{shapes.geometric, arrows, decorations.pathreplacing}

\newcommand{\xiren}{\xi_\text{ren}}
\newcommand{\xiinput}{\xi_\text{input}}
\newcommand{\betamatch}{\beta_\text{match}}

\newcommand{\dev}{\mathfrak{d}}

\newcommand{\comment}[1]{
\textcolor{black}{#1}}

\usepackage{hyperref}
\usepackage{cleveref}
\usepackage{authblk}

\title{Matching Lagrangian and Hamiltonian Simulations in (2+1)-dimensional U(1) Gauge Theory}

\author[1]{C.~F.~Groß \thanks{\href{mailto:gross@hiskp.uni-bonn.de}{gross@hiskp.uni-bonn.de}}}
\author[2]{S.~Romiti }
\author[1,3]{L.~Funcke}
\author[4,5]{K.~Jansen}
\author[6]{A.~Kan}
\author[5]{S.~Kühn}
\author[1]{C.~Urbach}

\affil[1]{Helmholtz Institute for Radiation and Nuclear Physics (HISKP) and Bethe Center for Theoretical Physics (bctp), Rheinische Friedrich-Wilhelms-Universität Bonn, Nußallee 14-16, 53115 Bonn, Germany}

\affil[2]{Institute for Theoretical Physics, Albert Einstein Center for
Fundamental Physics, University of Bern, CH-3012 Bern, Switzerland}

\affil[3]{Transdisciplinary Research Area ``Building Blocks of Matter and Fundamental Interactions'' (TRA Matter), Rheinische Friedrich-Wilhelms-Universität Bonn, Dechenstraße 3-11, 53115 Bonn, Germany}

\affil[4]{Computation-Based Science and Technology Research Center, The Cyprus Institute, 20 Kavafi Street, 2121 Nicosia, Cyprus}

\affil[5]{Deutsches Elektronen-Synchrotron DESY, Platanenallee 6, 15738 Zeuthen, Germany}

\affil[6]{Institute for Quantum Computing and Department of Physics \& Astronomy, University of Waterloo, Waterloo, Ontario, Canada, N2L 3G1}

\date{October 13, 2025}

\begin{document}
\maketitle

\begin{abstract}
    At finite lattice spacing, Lagrangian and Hamiltonian predictions differ due to discretization effects.
    In the Hamiltonian limit, i.e.\ at vanishing temporal lattice spacing $a_t$, the path integral approach in the Lagrangian formalism reproduces the results of the Hamiltonian theory.
    In this work, we numerically calculate the Hamiltonian limit of a $U(1)$ gauge theory in $(2+1)$ dimensions.
    This is achieved by Monte Carlo simulations in the Lagrangian formalism with lattices that are anisotropic in the time direction.
    For each ensemble, we determine the ratio between the temporal and spatial scale with the  static quark potential and extrapolate to $a_t \to 0$.
    Our results are compared with the data from Hamiltonian simulations at small volumes, showing agreement within $<2\sigma$.
    These results can be used to match the two formalisms.
\end{abstract}

\section{Introduction}
\label{sec:introduction}

Gauge theories are fundamental in our understanding of force mediation in the standard model (SM) of particle physics. 
Of the three forces unified in the SM, the strong force or quantum chromodynamics (QCD) is special because it is strongly coupled in the low energy regime.
Therefore, it requires a non-perturbative treatment, which is possible in the lattice regularisation of gauge theories pioneered by Wilson~\cite{Wilson:1974sk}.
While primarily applied to QCD, the lattice regularisation is applicable to any gauge theory.

In practice, any computation in a lattice gauge theory requires one to choose either the path integral formalism enabling mainly Monte Carlo (MC) simulations of such theories, or the Hamiltonian formalism.
The basis for the former has been provided already by Wilson, for the latter the corresponding Hamiltonian has been derived not much later based on general arguments in Ref.~\cite{Kogut:1974ag} by Kogut and Susskind, while Creutz derived the same expression for the Hamiltonian by explicitly constructing the transfer matrix~\cite{Creutz:1976ch}.
The Hamiltonian formulation has recently attracted fresh attention since it represents the natural formulation one would use on a future digital quantum computer.
Compared to the MC approach, Hamiltonian simulations have the advantage that for instance systems at finite density or real time evolution can be studied. 
However, the development state of current quantum computing devices limits such simulations to rather small systems.
Alternatively, tensor network states can be used, which see rapid development as well~\cite{Banuls:2018jag,Banuls:2019rao,Magnifico:2020bqt,Felser:2019xyv,Magnifico:2024eiy}.
Still, Hamiltonian simulations are currently restricted to systems with a relatively small number of degrees of freedom.

Ideas to nevertheless usefully apply Hamiltonian simulations already now include a clever combination with MC simulations, profiting from the respective strengths simultaneously~\cite{Clemente:2022cka,Crippa:2024cqr, Avkhadiev:2019niu, Avkhadiev:2022ttx, Avkhadiev:2024yjs,Carena:2021ltu,Carena:2022hpz}. 
One such idea has been brought forward in Ref.~\cite{Clemente:2022cka} and further investigated in Ref.~\cite{Crippa:2024cqr}.
It requires the matching of Hamiltonian and Lagrangian simulations: both formulations encompass the bare gauge coupling as a single parameter, which is directly related to the scale of the corresponding theory.
However, being a bare parameter implies that simply using the same numerical value for the coupling will, at least at finite lattice spacing, likely lead to sizeable artefacts.
An alternative is to connect the MC simulations with the Hamiltonian ones by taking the continuum limit in time direction, as the construction by Creutz suggests.
This continuum limit in time direction has been studied previously in Refs.~\cite{Loan:2002ej,Byrnes:2003gg,Loan:2003wy} based on the so-called anisotropic formulation of lattice gauge theories~\cite{Morningstar:1999rf}. Most relevant for our work here is Ref.~\cite{Loan:2002ej}, where the authors study a U$(1)$ theory using the anisotropic Wilson plaquette action. They take the temporal continuum limit keeping the $\beta$-value fixed. They compare to Green's Function Monte Carlo results which they comment, however, to be unreliable due to strong dependence on the trial wave function.

In this paper we will go beyond Refs.~\cite{Loan:2002ej,Byrnes:2003gg} in two ways: first, we will take the continuum limit in time direction in compact pure U$(1)$ gauge theory while keeping a suitable spatial length fixed, which we determine non-perturbatively. Second, we directly compare our extrapolated results from MC simulations with Hamiltonian simulations, finding agreement within $2 \sigma$.
It is conceptually straightforward to extend this to non-Abelian lattice gauge theories.
\comment{The continuum limit $a_s \to 0$ is not considered in this work, we rather want to compare the two formulations at fixed $a_s$.}
We reported on a first stage of this work in Ref.~\cite{Funcke:2022opx}.

The paper is organised as follows:
First, we give an overview of the theory in \cref{sec:theory-background}. 
Then we introduce the setup that we used for simulation in \cref{sec:setup}, and go into detail on our determination of the temporal continuum limit in \cref{sec:temp-contlimit}. 
We present our results in \cref{sec:results}, discuss them in \cref{sec:discussion}, and conclude in \cref{sec:outlook}.

\section{Theoretical background}
\label{sec:theory-background}

\subsection{Lagrangian formulation}
\label{subsec:theory-lagrangian}

On the Lagrangian side, we use the anisotropic Wilson action~\cite{Morningstar:1997ff,Morningstar:1999rf}, which reduces to the standard Wilson action for the special value of the anisotropy parameter $\xiinput=1$. It reads:
\begin{equation}
\label{eq:standard-wilson-action}
S_W =
\frac{\beta}{\xiinput}
\sum_{\vec{x}, i}
\operatorname{Re}
\left(1 - P_{0i}(\vec{x})\right)
+\beta \xiinput
\sum_{\vec{x}, i>j}
\operatorname{Re}
\left(1 - P_{i j}(\vec{x})\right)
,
\end{equation}
where $P_{\mu \nu}(\vec{x})$ is the so-called plaquette operator:
\begin{equation}
P_{\mu \nu}(\vec{x}) =
U_\mu(\vec{x}) \,
U_\nu(\vec{x}+\hat{\mu}) \,
U^\dagger_\mu(\vec{x}+\hat{\nu}) \,
U^\dagger_\nu(\vec{x})
\, .
\end{equation}
$\xiinput$ is the bare anisotropy and $\beta=1/g_0^2$ the inverse squared coupling constant.
The gauge links $U_\mu(\vec{x})$ are elements of U$(1)$ and can, hence, be parametrised as $U=e^{i\varphi}$ with a real-valued angle $\varphi$.
$\vec{x}$ is a point in our $2+1$ dimensional lattice and the directions $\mu \in \{0, 1, 2\}$.
MC simulations of the theory described by the action $S_W$ can be performed using standard Markov Chain MC methods, such as the Metropolis algorithm.
More details on the algorithm will be given below.

The renormalised anisotropy $\xi_\mathrm{ren}$ represents the ratio of temporal to spatial lattice spacing $a_t/a_s$. 
$\xi_\mathrm{ren}$ can be estimated from MC simulations in different ways, with the idea always being to compute one physical observable $O$ in units of both $a_t$ and $a_s$.
Once $a_t O$ and $a_s O$ have been determined, the renormalised anisotropy is estimated as
\begin{equation}
    \xi_\mathrm{ren}\ =\ \frac{a_t O}{a_s O}\ =\ \frac{a_t}{a_s}\,.
\end{equation}
In this paper we use two different choices for such an observable $O$, both based on the so-called static quark potential $V$, see for instance Ref.~\cite{Loan:2002ej}.
The static quark potential can be determined from planar Wilson loops $W(a_\mu x, a_\nu y)$ with extents $a_\mu x$ and $a_\nu y$.
Here, $a_\mu$ represents the lattice spacing in direction $\mu$, which can be one of the spatial directions or the time direction.
Since we are working in Euclidean space-time, the expectation values of Wilson loops decay exponentially in spatial as well as temporal directions.
By forming purely spatial Wilson loops $W_{ss}$ and spatial-temporal Wilson loops $W_{st}$, one obtains
\begin{equation}
\begin{split}
\lim_{y\to\infty}\frac{W_{ss}(x/a_s,(y+1)/a_s)}{W_{ss}(x/a_s,y/a_s)}&=\exp(-a_sV_s(x/a_s))\,,\\
\lim_{t\to\infty}\frac{W_{st}(x/a_s,(t+1)/a_t)}{W_{st}(x/a_s,t/a_t)}&=\exp(-a_tV_t(x/a_s))\,.
\end{split}
\label{eq:normalpotential}
\end{equation}
Again, due to the fact that we are working in Euclidean space-time, we have at equal distance $V_t = V_s$ up to a constant shift and, therefore, the anisotropy can be determined from a fit of 
\begin{align}
a_s V_s(x/a_s)&=\frac{1}{\xi_\mathrm{ren}}a_t V_t(x/a_s)+c\label{eq:fitnormalanisotropy}
\end{align}
to the data for the two potentials as a function of distance.
The fit parameter $c$ represents the difference in self-energy in $V_s$ and $V_t$ \comment{and arises from the fact that the potentials are measured along different axes. The self energy depends on the direction in which the quarks propagate~\cite{Klassen:1998ua}. In the above determination, the quarks propagate in the temporal direction in the measurement of the temporal potential, and propagate in the spatial direction in the measurement of the spatial potential.}
We refer to this procedure as the one based on the \enquote{normal} potential.

The second way to determine the potential is the one also used by Ref.~\cite{Alford:2000an}, and we refer to this method as the one based on the  \enquote{sideways} potential.
In this procedure the potential is determined varying the first argument in the corresponding Wilson loops as follows
\begin{equation}
\begin{split}
\lim_{x\to\infty}\frac{W_{ss}((x+1)/a_s,y/a_s)}{W_{ss}(x/a_s,y/a_s)}=\exp(-a_sV_s(y/a_s))\,,\\
\lim_{x\to\infty}\frac{W_{st}((x+1)/a_s,t/a_t)}{W_{st}(x/a_s,t/a_t)}=\exp(-a_sV_t(t/a_t))\,.
\end{split}
\label{eq:sidewayspotential}
\end{equation}
Now the argument is that if the potentials are equal, the distances in physical units must be equal as well. 
Thus, the anisotropy can be determined from 
\begin{equation}
V_s(y/a_s)\ =\ V_t\left(t/a_t\right)\quad\Rightarrow\quad y=t\quad\Rightarrow\quad \xi_\mathrm{ren} = \frac{a_t y}{a_s t} = \frac{a_t}{a_s}
\label{eq:determinationanisotropysideways}
\end{equation}
Since we rarely have spatial and temporal extents such that $V_s$ and $V_t$ are equal, we rescale the $y$-dependence of $V_s$ until the two curves $V_s(\xi_\mathrm{ren} y/a_s)$ and $V_t(t/a_t)$ agree within errors, which gives the renormalised anisotropy.

In practice, we start by interpolating linearly between any two neighbouring points $t/a_t$ and $(t+1)/a_t$ (excluding the smallest $t$-value) of the potential $V_t$. Next, we determine for each value of $V_s$ the corresponding scaling factor $\eta(y/a_s)$ by matching the value of $V_s(y/a_s)$ to the appropriate linear interpolation. Finally, we obtain $\xi_\mathrm{ren}$ by averaging over all $\eta(y/a_s)$.

\comment{Using the determinations of the Wilson loop above, the quarks propagate in the spatial direction in the measurement of both the spatial and temporal potential. Therefore the potentials have the same self energy effects.}

\subsubsection{Sommer parameter $r_0$ and setting the scale}

The static potential, which in $2+1$ dimensions has the form
\begin{equation}
    V(r)=a + \sigma r + d \ln(r)\,.
    \label{eq:potentialform}
\end{equation}
can be used to define a length scale $r_0$, the so-called Sommer parameter~\cite{Sommer:1993ce}, as follows
\begin{equation}
r^2\frac{\mathrm{d}}{\mathrm{d}r}V(r)|_{r=r_0}=c
\label{eq:rzerodefinition}
\end{equation}
in units of the spatial lattice spacing $a_s$. 
In QCD, the physical value of $r_0$ is known to be around $0.5\ \mathrm{fm}$ for $c=1.65$
, but, in the U$(1)$ theory at hand its physical value is unknown.
However, this is not relevant for our procedure, as we only need an observable in units of the spatial lattice spacing with a well defined continuum limit.
We use a value of $c=1.65$, because it turns out to be in the linear region of the potential and we stick to the notation $r_0/a_s$. 
Once the potentials are parametrised using the form \cref{eq:potentialform}, the corresponding value of $r_0/a_s$ can be determined.
It allows one to fix the spatial lattice spacing: if in two simulations with parameters $(\beta, \xi)$ and $(\beta',\xi')$ the same $r_0/a_s$ is measured within uncertainties, both simulations exhibit the same $a_s$.

In practice, we determine $r_0$ by fitting the potential form \cref{eq:potentialform} to the data. Then, we take the derivative analytically and express $r_0/a_s$ as follows 
\begin{equation}
    \frac{r_0}{a_s} = -\frac{d}{2\sigma} + \sqrt{ \left( \frac{d}{2\sigma}\right)^2-\frac{c}{\sigma}}\,,
\end{equation}
with the parameters $\sigma$ and $d$ from \cref{eq:potentialform} and $c$ from \cref{eq:rzerodefinition}.

\subsection{Hamiltonian}
\label{subsec:theory-hamiltonian}

The Kogut-Susskind Hamiltonian for the pure U(1) lattice gauge theory in (2+1) dimensions is given by~\cite{Kogut:1974ag}
\begin{equation}\label{eq:H}
\hat{H}_\text{tot} = \frac{g^{2}}{2} \sum_{\vec{r}}\left(\hat{E}^{2}_{\vec{r}, 1} 
	+ \hat{E}^{2}_{\vec{r}, 2}\right) -\frac{1}{2a^2g^{2}} \sum_{\vec{r}} \left(\hat{P}_{\vec{r}} + \hat{P}_{\vec{r}}^{\dag}
    \right)\,,
\end{equation}
where $a$ is the lattice spacing and $g$ is the bare coupling. The operator $\hat{E}_{\vec{r}, \mu}$ represents the dimensionless electric field on the link starting from the lattice site at coordinates $\vec{r}=(r_1, r_2)$ in the direction $\mu \in \{1, 2\}$. The plaquette operator $\hat{P}_{\vec{r}} =  \hat{U}_{\vec{r},1}\hat{U}_{\vec{r}+1,2}\hat{U}^{\dag}_{\vec{r}+2,1}\hat{U}^{\dag}_{\vec{r},2}$  is defined as the product of four unitary link operators $\hat{U}_{\vec{r},\mu}$, where the notation $\vec{r}+1 \equiv (r_1+1, r_2)$ and $\vec{r}+2 \equiv (r_1, r_2+1)$ indicates the neighbouring sites in the $1$ and $2$ directions, respectively. The link operator is defined as
\begin{equation}
\hat{U}_{\vec{r},\mu}=e^{iag\vec{A}_{\vec{r},\mu}}\,,
\end{equation}
where $\vec{A}_{\vec{r},\mu}$ is the discretized vector field in the compact formulation, i.e., the values of $ag\vec{A}_{\vec{r},\mu}$ are constrained to lie within the interval $[0, 2\pi)$. Note that $\hat{U}_{\vec{r},\mu}$ is a unitary operator. The commutation relations between the electric field operator $\hat{E}_{\vec{r},\nu}$ and the link operator $\hat{U}_{\vec{r'},\mu}$ read
\begin{align}
    [\hat{E}_{\vec{r},\nu},\hat{U}_{\vec{r'},\mu}]&=\delta_{\vec{r},\vec{r'}}\delta_{\nu,\mu} \hat{U}_{\vec{r},\nu},\label{eq:U_E_commutation_relation1}\\
    [\hat{E}_{\vec{r},\nu},\hat{U}^{\dag}_{\vec{r'},\mu}]&=-\delta_{\vec{r},\vec{r'}}\delta_{\nu,\mu}\hat{U}^{\dag}_{\vec{r'},\nu}.
    \label{eq:U_E_commutation_relation2}
\end{align}
The gauge-invariant states satisfy Gauss's law at every site $\vec{r}$, 
\begin{align}\label{eq:gauss}
    \Bigg[\sum\limits_{\mu=1, 2}
\left(\hat{E}_{\vec{r}, \mu} -\hat{E}_{\vec{r}- \mu, \mu} \right) &- Q_{\vec{r}}\Bigg] \ket{\Phi} = 0,
\end{align}
where $Q_{\vec{r}}$ are the static charges. 
Instead of considering the full Hilbert space and enforcing Gauss's law, we formulate the theory directly on the gauge-invariant subspace, by using the Gauss's law constraints to eliminate certain degrees of freedom~\cite{Haase:2020kaj,Kaplan:2018vnj,Paulson:2020zjd,Kan:2021nyu,Bauer:2021gek}.

For a numerical implementation of the Hamiltonian, the gauge degrees of freedom must be truncated to a finite dimension because the electric field values are unbounded, resulting in an infinite-dimensional Hilbert space for these degrees of freedom. The continuous $U(1)$ gauge group can be discretized in the electric basis to the group of integers $\mathbb{Z}_{2l+1}$. The integer $l$ sets the truncation level, i.e., the discretized gauge fields are constrained to integer values within the range $[-l, l]$~\cite{Haase:2020kaj}. The total Hilbert space dimension is $(2l+1)^N$, where $N$ is the number of gauge fields.

The eigenstates $e_{\vec{r}, \mu}$ of the electric field operator $\hat{E}_{\vec{r}, \mu}$ form a basis,
\begin{equation}
    \hat{E}_{\vec{r}, \mu} \ket{e_{\vec{r}, \mu}}=e_{\vec{r}, \mu}\ket{e_{\vec{r}, \mu}} , \ \ \ e_{\vec{r}, \mu} \in [-l,l]
    \, ,
\end{equation}
on which the link operators $\hat{U}_{\vec{r},\mu}$ and $\hat{U}^{\dag}_{\vec{r},\mu}$ act as raising and lowering operators, respectively,
\begin{equation}
    \hat{U}_{\vec{r},\mu} \ket{e_{\vec{r}, \mu}}=\ket{e_{\vec{r}, \mu}+1},\ \ \  \hat{U}^{\dag}_{\vec{r},\mu}\ket{e_{\vec{r}, \mu}}=\ket{e_{\vec{r}, \mu}-1}.
\end{equation}
When discretizing a gauge theory, one needs to give up either unitarity or the exact commutation relations between the electric field and link operators in \cref{eq:U_E_commutation_relation1} and \cref{eq:U_E_commutation_relation2}. In our case, the commutation relations are preserved for the truncated operators, but unitarity is lost, $\hat{U}^{\dag}_{\vec{r},\mu}\hat{U}_{\vec{r},\mu}\neq \mathbbm{1}$. This can be seen from the matrix representation of the link operators~\cite{Paulson:2020zjd},
\begin{gather}\label{eq:uoperator}
\hat{U} \mapsto
\begin{pmatrix}
  0 &  \dots & \dots & 0 \\
  1 &  \dots & \dots & 0 \\
  0 &  \ddots & \vdots & 0 \\
  0 &  \dots & 1 & 0\\
\end{pmatrix},\ \ \ \hat{U}^{\dag} \mapsto
\begin{pmatrix}
  0 &  1 & \dots & 0 \\
  0 &  \vdots & \ddots & 0 \\
  0 &  \dots & \dots & 1 \\
  0 &  \dots & \dots & 0\\
\end{pmatrix}.
\end{gather}
However, unitarity is recovered in the limit of $l\to \infty$. The resulting errors due to the finite truncation parameter $l$ have been investigated in Refs.~\cite{Kuhn:2014rha,Buyens:2017crb}. Alternative methods for defining the electric field and link operators have been explored in Refs.~\cite{Mathis:2020fuo,Chandrasekharan:1996ih,Wiese:2013uua,Hashizume:2021qbb}.

In order to analytically derive the Kogut-Susskind Hamiltonian in \cref{eq:H} from the Wilson action in \cref{eq:standard-wilson-action}, one needs to employ the transfer matrix method~\cite{Creutz:1976ch}. This derivation has been performed for various quantum field theories in arbitrary dimensions, for example, for studying transport coefficients~\cite{Cohen:2021imf} and the topological $\theta$-term of (non)Abelian lattice gauge theories in (3+1) dimensions~\cite{Kan:2021nyu}. Since the Wilson action is defined on a $(d+1)$-dimensional space-time lattice and the Kogut-Susskind Hamiltonian is defined on a $d$-dimensional spatial lattice, the limit of $a_t\to 0$ has to be taken when deriving \cref{eq:H} from \cref{eq:standard-wilson-action} using the transfer matrix formalism~\cite{Creutz:1976ch}. 

The parameters of the resulting Kogut-Susskind Hamiltonian generally differ from the original parameters of the Wilson action, due to renormalization effects. Thus, when combining Hamiltonian and Lagrangian lattice methods, these parameters need to be matched.

\section{Methods}
\label{sec:setup}

\subsection{Lagrangian}
\label{subsec:setup-lagrangian}

We use two different Markov Chain Monte Carlo algorithms to simulate the lattice action \cref{eq:standard-wilson-action}.
We use periodic boundary conditions in all directions.
For values of $\xiinput\geq 1/4$ we use the standard Metropolis algorithm, where
each link is updated 10 times per sweep.
We discard an appropriate amount of sweeps to account for thermalisation, and only analyse every 50th or every 100th configuration to account for autocorrelation effects.
For anisotropies smaller than $\xiinput=1/4$, we encounter issues with critical slowing down and,
therefore, use a combination of heatbath and overrelaxation algorithms, with ten heatbath steps per overrelaxation step. Only in the case of $\xiinput=0.18$, five heatbath steps are followed by five overrelaxation steps.

Details of the algorithm can be found in Ref.~\cite{Hattori:1992qk} (see in particular the arXiv version).

\comment{We provide our codes for the generation of the gauge ensembles and for the analysis of the ensembles in~\cite{FK2/AUJJ08_2025}\footnote{\comment{Please see the README-files for more information on how to generate the ensembles and how to replicate the analysis.}}. For the analysis, we have used the library~\cite{kostrzewa_2024_14800686}.}

When performing the limit $\xi\to0$, we keep the spatial volume $(a_sL)^2$ fixed and scale the time extent $T$ by $\xiinput^{-1}$ in order to keep both the physical spatial and time extents constant.

In total, we have generated 82 ensembles with $L=16$ with $\beta$-values in the range $(1.39, 1.75)$ and $\xiinput$-values $1, 4/5, 2/3, 1/2, 2/5, 1/3, 1/4, 1/5, 0.18$. 
Even smaller values of the anisotropy turned out to be unrealistic due to too long equilibration and autocorrelation times.
A list of all ensembles is compiled in the appendix in \cref{tab:usedconfigsL16} together with relevant parameter values, algorithm, autocorrelation times, and number of configurations included in the analysis.
The bootstrap method is used for the statistical analysis with 500 bootstrap samples. Residual autocorrelation times are taken into account as discussed in \cref{sec:autocorrelation}.

In principle, one would perform the stochastic simulations in $2+1$ dimensions in the very same spatial volume used also in the Hamiltonian simulations. However, this target volume is so small that the static potential in the relevant region of distances, $r_0/a_s$ and the renormalised anisotropy cannot be determined reliably.
Therefore, a two-step procedure is required in which $r_0/a_s$ and the parameter-values for the temporal continuum limit are determined in large spatial volume. This is followed by dedicated simulations with $L=3$ to match the volume of the Hamiltonian simulations.

In addition to the $L=16$ simulations, we have also generated 54 dedicated $L=3$ ensembles, which exactly match the spatial volume used in the Hamiltonian simulations. These are listed in the appendix in \cref{tab:usedconfigsL3}. The time extent $T$ for given $\xiinput$ was chosen equal to the large volume simulations at the same $\xiinput$.

Further, the range of $\beta$-values we can use in practice for the matching is restricted: for too small $\beta$-values $a_s$ is too large to reliably determine the static potential from Lagrangian simulations.
On the other hand, for too large $\beta$-values, the Hamiltonian simulations we are using are suffering from significant truncation errors.
This leaves us currently with a window of $\beta$-values in the limit $\xi\to0$ between $\beta=1.35$ and $\beta=1.5$.

\subsubsection{Determining $r_0$ and $\xiren$}

Once we have determined the Wilson loops, we extract the potential by computing the ratios
\cref{eq:normalpotential,eq:sidewayspotential}, and determine the values of the potential using fits to effective masses. In order to account for ambiguities in the choice of the fit range,
we perform a model averaging procedure, see \cref{sec:aic}. This allows us to compute statistical or combined statistical and systematic errors for each value of the potential.
From the potential we determine $\xiren$ and $r_0/a_s$ from different ranges of distances in the intervals $I_{\xiren}$ and $I_{r_0}$, respectively.
The various choices can be combined in different ways, which we use to define analysis chains, all of which are compiled in \cref{tab:analysischains}.
The most important difference between the analysis chains is whether they include a systematic error (label ET) or not (label ES).

\begin{table}[]
    \centering
    \begin{tabular}{lllll}
    \hline
    label & Potential type & $I_{\xiren}$ & $I_{r_0}$& included error\\
    \hline\hline
N1ES & Normal   & $[2, 7]$ & $[1, 7]$ & statistical only\\
S1ES & Sideways & $[2, 7]$ & $[1, 7]$ & statistical only\\
N0ES & Normal   & $[2, 8]$ & $[1, 8]$ & statistical only\\
S0ES & Sideways & $[2, 8]$ & $[1, 8]$ & statistical only\\
N1ET & Normal   & $[2, 7]$ & $[1, 7]$ & statistical and systematic\\
S1ET & Sideways & $[2, 7]$ & $[1, 7]$ & statistical and systematic\\
N0ET & Normal   & $[2, 8]$ & $[1, 8]$ & statistical and systematic\\
S0ET & Sideways & $[2, 8]$ & $[1, 8]$ & statistical and systematic\\
\hline
    \end{tabular}
    \caption{Summary of the different analysis chains.
The potential type refers to those defined in \cref{subsec:theory-lagrangian}. The intervals $I_{\xiren}$ and $I_{r_0}$ indicate the range of distances in units of the spatial lattice spacing $a_s$ used to determine $\xiren$ and $r_0/a_s$, respectively.
For the different error determinations see \cref{sec:aic}.
The systematic error arises from uncertainty in choosing the fit range for the effective masses.
In the following, we refer to the different analysis chains with the labels in the first column (see text).
}
    \label{tab:analysischains}
\end{table}

\subsubsection{Taking the temporal continuum limit}
\label{sec:temp-contlimit}

For the generation of ensembles we start with isotropic simulations corresponding to $\xi=\xiinput=1$ at a given $\beta$-value $\beta=\beta_\mathrm{iso}$ and determine $r_0/a_s(\beta_\mathrm{iso})\equiv r_\mathrm{iso}$.
Then, for all $\xiinput$-values smaller than one, we perform simulations for several $\beta$-values in the region of $r_0/a_s$-values close to $r_\mathrm{iso}$, until one ensemble reproduces $r_\mathrm{iso}$ within errors.
We denote the corresponding $\beta$-parameter with $\beta=\betamatch$ and the plaquette-value with $P=P_\text{match}$, which are the ones we use for the continuum extrapolation. Note that $\betamatch$ and $P_\mathrm{match}$ depend on the particular analysis chain.

By definition, the value of $\beta_\mathrm{match}$ has no statistical uncertainty, because it is the input value to the simulation of the corresponding matching ensemble.
This is unrealistic and, as it turns out, also impractical for the remaining analysis.
Thus, we perform a linear bootstrap fit to the data of $r_0/a_s$ as function of $\beta$ at fixed $\xiinput$, and use the bootstrap error from the fit as an estimate for the statistical uncertainty of $\beta_\mathrm{match}$. In the fit we only include ensembles with $r_0/a_s$ differing from $r_\mathrm{iso}$ by less than $0.3$.
$\xiren$ is taken directly from the matching ensemble, and it is also determined separately for each analysis chain.
$P_\mathrm{match}$, on the other hand, has only statistical uncertainties.

\begin{figure}
    \centering
    \includegraphics[width=\linewidth]{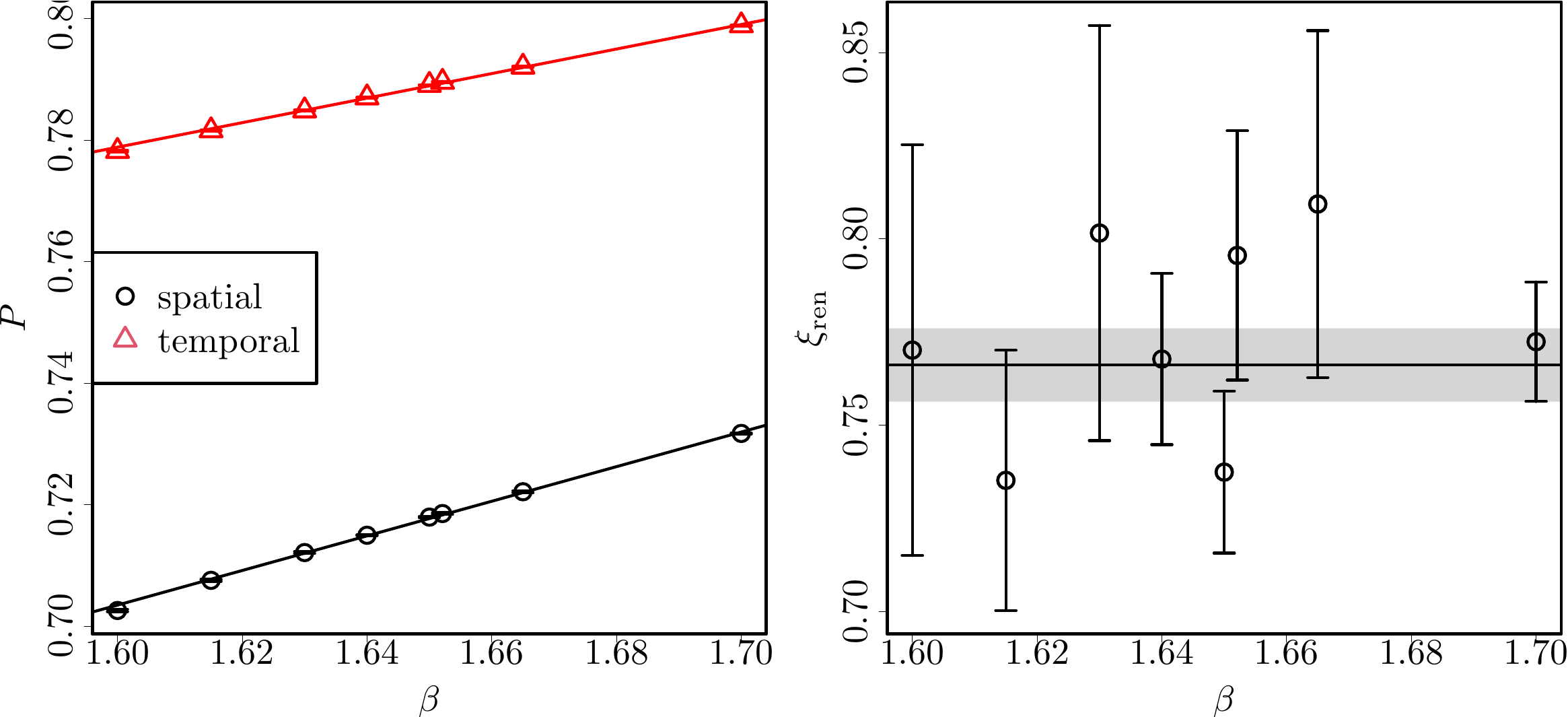}
    \caption{Spatial and temporal plaquette (left panel) and renormalised anisotropy (right panel) as functions of $\beta$ for $\xiinput=0.8$ and $\beta_\text{iso}=1.7$. $\xiren$ is determined from the analysis chain N0ET, see \cref{tab:analysischains}.
    The solid lines represent fits to the data, and the shaded regions the corresponding bootstrap errors.
    The fits are linear in $\beta$ for the plaquette, and a constant for $\xiren$.
    }
    \label{fig:pofbetaxiofbeta}
\end{figure}

In principle, there is also a systematic effect from choosing the matching ensemble: there might be two simulation points with $r_0/a_s$-values equally close within errors to $r_\mathrm{iso}$.
However, in practice this appeared to be irrelevant, since in particular $\xiren$ is basically independent of $\beta$ at fixed $\xiinput$.

In \cref{fig:pofbetaxiofbeta} we show the spatial-spatial and spatial-temporal plaquette as a function of $\beta$ on the left-hand side and the renormalized anisotropy $\xiren$ as a function of $\beta$ on the right-hand side, both at fixed $\xiinput=0.8$, with $\beta_\text{iso}=1.7$ and $L=16$.
We see that the plaquette behaves linearly with $\beta$, whereas $\xiren$ does not depend on $\beta$ within errors.

In selecting the matching $\beta$ for every $\xiinput$, we can define a trajectory of constant spatial lattice spacing for each analysis chain.

For every point along these trajectories
we then perform dedicated $L=3$ simulations to control finite volume effects. We apply two different procedures for the finite size correction: either we directly extrapolate the small volume results to the continuum limit, or we first extrapolate the ratio
\begin{equation}
\label{eq:ratio}
    R(\xiren^2) = \frac{P(L=3, \xiren^2)}{P(L=16, \xiren^2)}
\end{equation}
and combine them with the extrapolated value of $P(L=16)$ to correct for finite size effects in the continuum limit.
We call the two methods \enquote{direct} and \enquote{ratio}, respectively.

The temporal continuum extrapolation of the spatial plaquette and $\beta$ is then performed by fitting polynomials $\mathcal{P}_{n_p}(\xiren^2)$ of degree $n_p$ in $\xiren^2$ to the data, equivalent to an extrapolation in $a_t^2$, which is expected for a pure gauge theory.
The extrapolations are performed with different fit ranges in $\xiren$, and with different degrees $n_p$.
In particular, we have two sets of continuum extrapolations denoted as cA and cB, respectively, which mainly differ by the inclusion or exclusion of $\xiinput=0.18$ in cA and cB, respectively.
The two different sets of fits are compiled in \cref{tab:groupcontlimit}.

\begin{table}[tbp]
    \centering
    \begin{tabular}{l|l|l|l}
    	            \multicolumn{2}{c|}{cA}             &             \multicolumn{2}{c}{cB}             \\
    	\hline
    	$\xiinput$-values                    & $n_p$ & $\xiinput$-values                   & $n_p$ \\
    	\hline
    	$0.18, 1/5, 1/4$                & 1               & $1/5, 1/4, 1/3$                & 1               \\
    	$0.18, 1/5, 1/4, 1/3$           & 1               & $1/5, 1/4, 1/3, 2/5$           & 1               \\
    	$0.18, 1/5, 1/4, 1/3, 2/5$      & 2               & $1/5, 1/4, 1/3, 2/5, 1/2$      & 2               \\
    	$0.18, 1/5, 1/4, 1/3, 2/5, 1/2$ & 2               & $1/5, 1/4, 1/3, 2/5, 1/2, 2/3$ & 2
    \end{tabular}
    \caption{Two sets of extrapolations cA and cB used to calculate a combined continuum limit of the fits. The same fit ranges are used for all trajectories and for all fits to the continuum limit -- for $\beta$, the plaquettes and for $R$.
    We list the anisotropies that are included in the polynomial fits and the degrees of the polynomials.}
    \label{tab:groupcontlimit}
\end{table}

For each of the eight analysis chains, we perform all eight fits listed in \cref{tab:groupcontlimit}, leading to 64 continuum limits, for the pair $(\beta, P)$.

The 64 pairs fall into four sets of equal size corresponding to the extrapolation set--analysis chain combinations (cA-ES), (cB-ES), (cA-ET) and (cB-ET), respectively.

We extract the statistical error of the final results of the sets cA and cB from the standard deviation of the bootstrap distribution of the unweighted average over all pairs in (cA-ES) and (cB-ES), respectively.
Likewise, the combination of statistical and systematic error is obtained from (cA-ET) and (cB-ET).
Additionally, we include for (cA-ET) and (cB-ET) separately the spread of the different continuum results as follows
\begin{equation}
\sigma_\text{spread, tot}^2 = \sigma_\text{unweighted}^2 + \frac{1}{N} \sum_i \left(\mu_\text{unweighted} - \mu_i\right)^2
\label{eq:syserrweightedaverage}
\end{equation}
in the corresponding systematic uncertainty. Here, $(\mu,\sigma)_\text{unweighted}$ represents the mean and standard deviation from the combined bootstrap samples, and $\mu_i$ the mean values of the single fits.
We fold this systematic uncertainty into our bootstrap distribution of (cA-ET) and (cB-ET), respectively, by an appropriate rescaling, analogous to what is described in \cref{sec:autocorrelation}.
A flow chart of this procedure is given in \cref{fig:flowchartanalysis_extended}.

To summarise this technical discussion:
the procedure described above leaves us with a purely statistical error $\sigma_\text{stat}$, a combined error $\sigma_\text{comb}$ from the statistical error and the systematic errors from choosing the plateau points, and a combination of $\sigma_\text{comb}$ and the error due to the spread, $\sigma_\text{spread,tot}$.
In the end we can isolate the single errors using the relations $\sigma_\text{spread}^2=\sigma_\text{spread,tot}^2 - \sigma_\text{comb}^2$ and $\sigma_\text{pot}^2 = \sigma_\text{comb}^2 - \sigma_\text{stat}^2$ and eventually quote the errors $\sigma_\text{spread}, \sigma_\text{pot}$ and $\sigma_\text{stat}$.
As final temporal continuum results for the observables $\beta, P(L=16), P(L=3)$ and $R$, we quote the mean values from the unweighted averages over (cA-ET) and (cB-ET).

\subsection{Hamiltonian}
\label{subsec:setup-hamiltonian}

As discussed in section 2.2, we use the Gauss's law to eliminate some gauge degrees of freedom, thereby restricting the theory to the gauge-invariant space.
More specifically, we treat the Gauss's law in \cref{eq:gauss} as a set of constraints on the electric operators, and solve this set of equations over the electric operators.
While there are $N$ Gauss's law constraints, they are not independent, since there is a conservation of charges, which, in the pure gauge case, means the constraints sum to zero.
Therefore, there are only $N-1$ independent constraints, which allows us to express $N-1$ arbitrary electric field operators, i.e., effectively eliminating them, in terms of the remaining ones.
Since the eliminated electric fields do not contribute directly to the dynamics, their corresponding link operators become identities.
For a two-dimensional $L \times L$ square lattice, where $N=L^2$, $L^2-1$ out of the $2L^2$ gauge degrees of freedom are eliminated.
Thus, the Hamiltonian is expressed in terms of the $L^2 + 1$ remaining gauge fields. 
This reduces the number of basis states from $(2l+1)^{2L^2}$ to $(2l+1)^{L^2+1}$, which in practice, alleviates the computational costs significantly.

Here we perform exact diagonalization of a $3\times 3$ lattice with periodic boundary conditions to solve for the ground state $| \Psi_0\rangle$.
Then, we evaluate the plaquette expectation value, defined by
\begin{equation}
    \langle \mathcal{P} \rangle \equiv \Bigg\langle \Psi_0\Bigg| \frac{1}{2V}\sum_{\vec{r}} \left(\hat{P}_{\vec{r}} + \hat{P}^\dag_{\vec{r}}\right) \Bigg| \Psi_0\Bigg\rangle,
\end{equation}
where $V$ is the number of plaquettes in the lattice. Note that we set the lattice spacing $a=1$ throughout our simulations.
\comment{Truncating the electric field on the dynamical links to a range of $[-l,l]$ can in principle lead to configurations where the links that have been eliminated implicitly carry an electric flux exceeding this range. These would violate discrete symmetries of the Hamiltonian and are unphysical. To ensure that our results are not affected by such effects,} the simulations are carried out over a range of $1/g^2$ in $(0,10]$ and for $l \in \{ 1, 2, 3, 4 \}$. We find that the values of $\langle \mathcal{P} \rangle$ obtained for $l=2,3,4$ agree with each other up to $1/g^2 = 1.5$, \comment{and are thus not affected by truncation artifacts. Beyond these values of the inverse coupling,} the values start to deviate for different $l$ values, indicating that the simulations are no longer reliable.

\subsection{Comparing Lagrangian and Hamiltonian simulations}
\label{sec:compareLandH}

We compare the Lagrangian results, obtained with the methods in \cref{subsec:setup-lagrangian} and the Hamiltonian results, obtained with the methods in \cref{subsec:setup-hamiltonian}, in the two-dimensional $\beta_\text{match}$-$P_\text{match}$-plane.
We use confidence ellipses in addition to error bars to display the errors of our measurements.
The confidence ellipses are constructed from the errors of the continuum limit results of $\beta_\text{match}$ and $P_\text{match}$ and the correlation between them.
To quantify the deviation between Lagrangian and Hamiltonian results, we scale the confidence ellipse until it is touching the interpolation of the Hamiltonian result.
Then we convert the radius of this touching ellipse into the probability that a point from the Lagrangian distribution is within the ellipse and thus not a match to the Hamiltonian, i.e.\ the probability that the Hamiltonian and Lagrangian measurements do not match.
We convert the mismatch probability into the probabilities of standard deviations of the normal distribution.
The detailed formula for the ellipses and the probabilities are given in \cref{sec:confidencelevel}.

\section{Results}
\label{sec:results}

In this section, we mainly present results of the stochastic simulations. The results from the Hamiltonian simulations are only required at the end when we compare it with the results in the temporal continuum limit.

\begin{figure}[t]
    \centering
    \includegraphics[width=\linewidth]{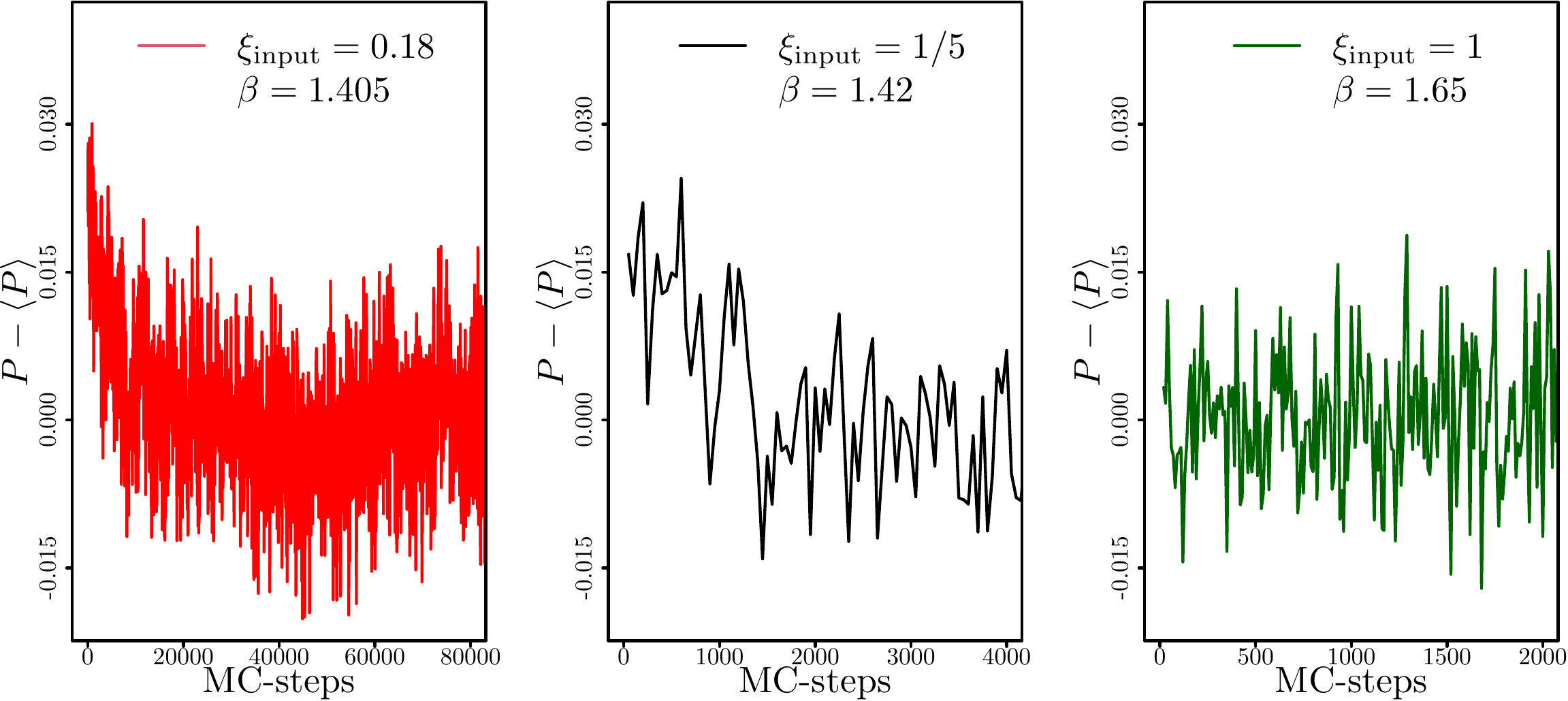}
    \caption{Thermalisation of the plaquette for the input anisotropies $\xiinput \in\{0.18, 1/5, 1\}$. All data were generated by the heatbath-overrelaxation algorithm, and the ensembles have the lattice sizes $L^2\times T = 16^2 \times 88, 80, 16\approx16^2 \times 16/\xiinput$. We show the difference between the measurement and the mean value of the plaquette.}
    \label{fig:thermalizationtriple}
\end{figure}

In \cref{fig:thermalizationtriple}, we exemplarily show three equilibration histories of the plaquette: we plot $P-\langle P\rangle$ as a function of the number of Monte Carlo steps, where $\langle P\rangle$ is computed after equilibrium is reached.
Each step corresponds to a complete sweep over the lattice. The main difference between the three panels is the value of $\xiinput$, which is $\xiinput=0.18$ in the leftmost panel, $\xiinput=0.2$ in the middle, and $\xiinput=1$ in the rightmost panel.
The ensembles have constant $a_s$, leading to different $\beta$. The values of $\beta_\text{match}$ were taken from the analysis chain N0ET and $\beta_\text{iso}=1.65$.
For $\xiinput=1$, thermalisation is achieved almost instantly, for $\xiinput=1/5$ it takes about $2000$ Monte Carlo steps, and for $\xiinput=0.18$, thermalisation is only achieved after about $25000$ steps. Also, one observes long-range fluctuations at the smallest value of the anisotropy, hinting at larger autocorrelation times.

\begin{figure}[htbp]
    \centering
    \includegraphics[width=0.7\linewidth]{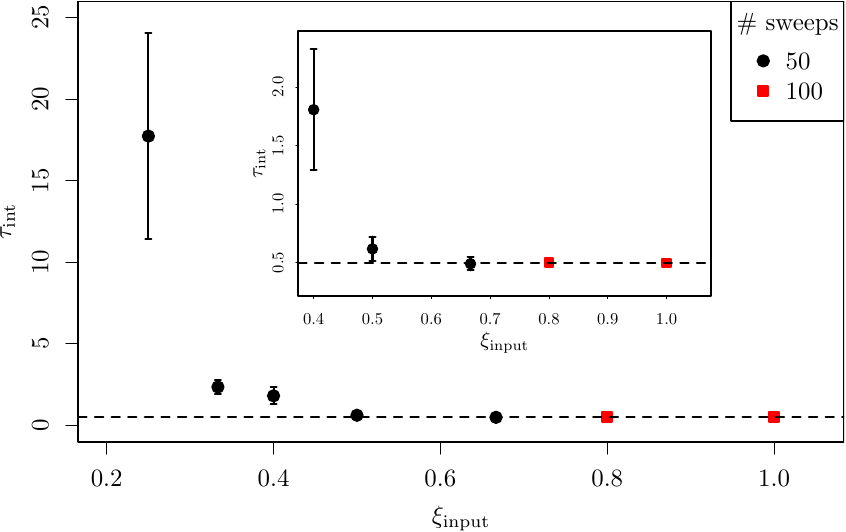}
    \caption{Integrated autocorrelation time of the plaquette for different input anisotropies. $\beta=1.7$ is kept constant, the red squares correspond to points with 100 sweeps between measurements, the black circles to 50 sweeps between measurements.
    The dashed line shows the ideal case $\tau_\text{int}=0.5$.
    The inset is a close-up of the lower right region of the larger figure.
    All simulations were done with the Metropolis-algorithm.
    }
    \label{fig:autocorrelation}
\end{figure}

In the following, we show results exemplarily for the analysis chain N0ET.
We show the integrated autocorrelation times for constant $\beta$ and simulation with the Metropolis-algorithm in \cref{fig:autocorrelation}, and we observe that the autocorrelation increases roughly exponentially with decreasing $\xiinput$.
For anisotropies that are close to 1, we see an ideal autocorrelation time with $\tau_\text{int} \approx 0.5$.
For decreasing $\xiinput$, the error on the integrated autocorrelation time grows larger, but also $\tau_\text{int}$ itself grows.
For $\xiinput=1/4$, the autocorrelation grows to $\tau_\text{int}=17.7(63)$.
This prevents us from simulating even smaller anisotropies with the Metropolis-algorithm.

\begin{figure}
    \centering
    \includegraphics[width=0.7\linewidth]{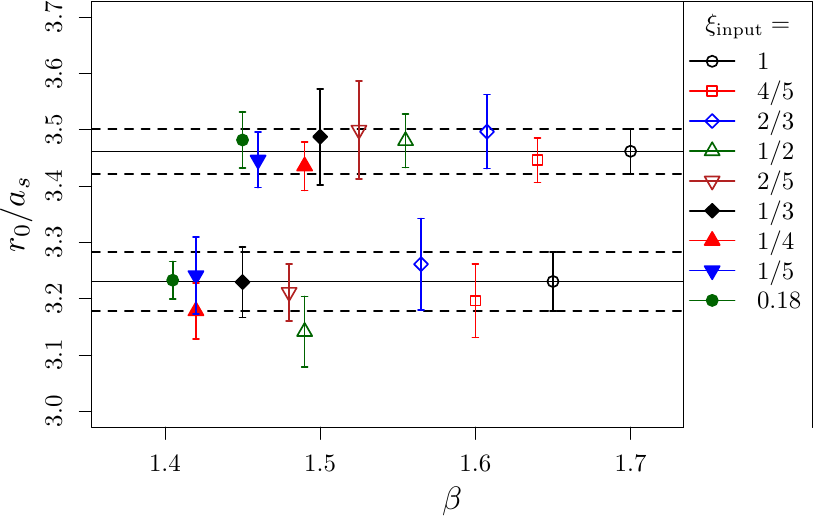}
    \caption{The matching points for analysis chain N0ET.
    The upper points correspond to the matching points for $\beta_\mathrm{iso}=1.7$, the lower ones to $\beta_\mathrm{iso}=1.65$.}
    \label{fig:allmatchingpoints}
\end{figure}

In \cref{fig:allmatchingpoints}, we illustrate the tuning procedure to determine $\beta_\mathrm{match}$ at fixed $r_0/a_s=r_\mathrm{iso}$ for two values of $\beta_\mathrm{iso}=1.65$ and $\beta_\mathrm{iso}=1.7$. We plot $r_0/a_s$ of the matching ensemble as a function of $\beta$, where the different symbols indicate different values of $\xiinput$, as indicated in the figure legend. The value of $r_\mathrm{iso}$ determined at $\beta_\mathrm{iso}$ is shown as the black circle with statistical uncertainty, and -- to guide the eye -- also by the dashed horizontal line with error band.

\begin{figure}
    \centering
    \includegraphics[width=0.5\linewidth]{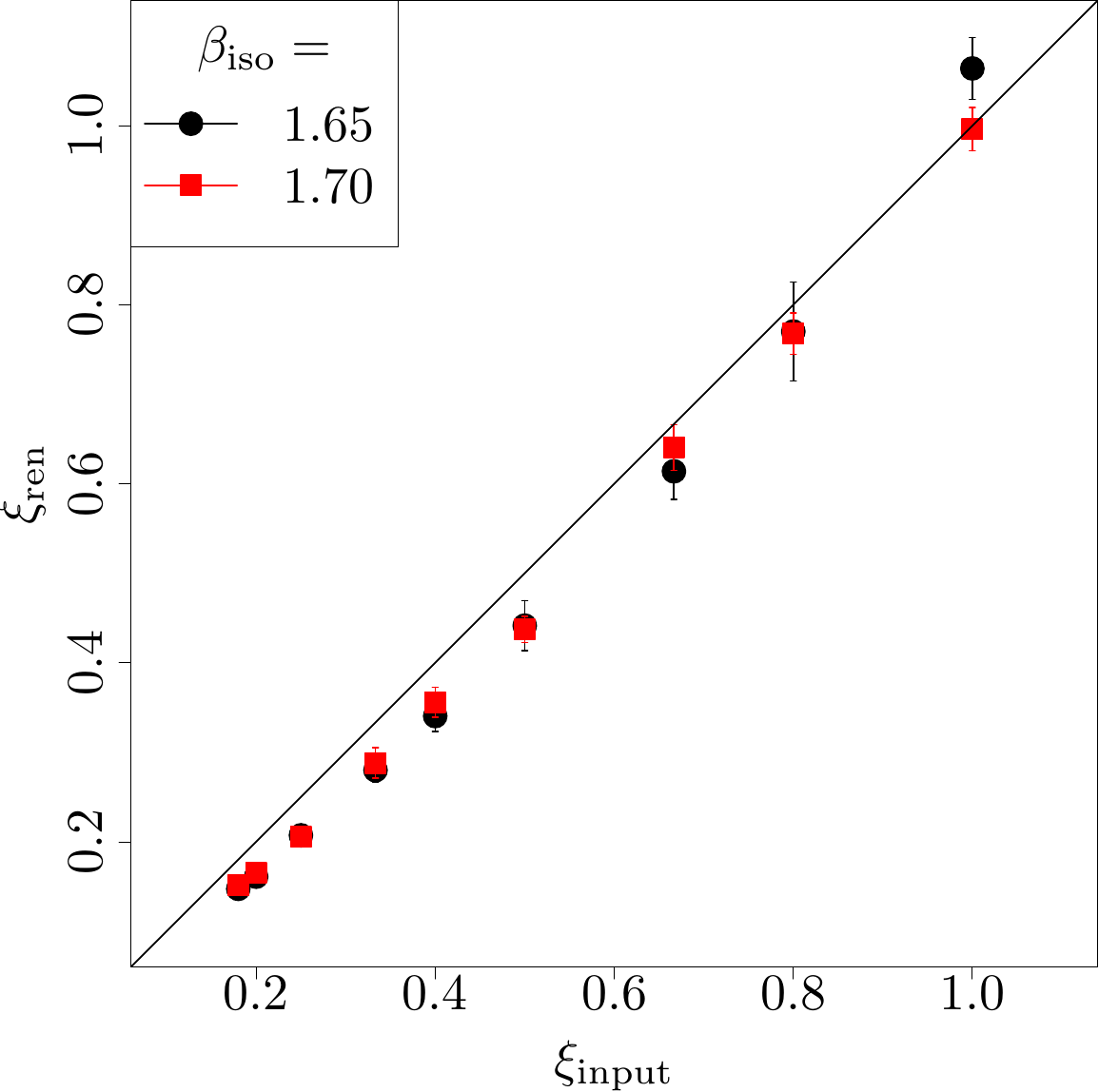}
    \caption{$\xiren$ at the matching points for analysis chain N0ET.
    The red squares correspond to the matching points for $\beta_\mathrm{iso}=1.7$, the black circles to $\beta_\mathrm{iso}=1.65$.
    The diagonal line shows the line of $\xiren=\xiinput$.
    }
    \label{fig:anisotropyatmatching}
\end{figure}

In \cref{fig:anisotropyatmatching}, we show the renormalized anisotropy as a function of the input anisotropy at the matching points, i.e.\ at fixed lattice spacing, for the same values of $\beta_\text{iso}$.
The diagonal line shows the line of $\xiinput=\xiren$, and we see that in most cases, $\xiren$ is smaller than $\xiinput$.
However the deviation is small, the maximum deviation is $25\%$, and the median deviation is $15\%$.
We observe only small differences in $\xiren$ when comparing the two $\beta_\text{iso}$-values.

For convenience, we compile the results
in \cref{tab:res1.7} for $\beta_\mathrm{iso}=1.7$ and in \cref{tab:res1.65} for $\beta_\mathrm{iso}=1.65$. For all our $\xi_\mathrm{input}$ values we list the corresponding values of $\xi_\mathrm{ren}$, $P_\mathrm{match}$, and $\beta_\mathrm{match}$. These results are again exemplary for all the different analysis chains.

\begin{table}[]
    \centering
    \begin{tabular}{lrrrr}
    \hline
    $\xi_\mathrm{input}$ & $\beta_\mathrm{match}$ & $\xiren$ & $P_\mathrm{match}(L=16)$ & $P_\mathrm{match}(L=3)$\\
    \hline$4/5$ & $1.6400(79)$ & $0.768(23)$ & $0.714953(69)$ & $0.73809(39)$ \\
$2/3$ & $1.6075(82)$ & $0.641(26)$ & $0.69608(16)$ & $0.71914(38)$ \\
$1/2$ & $1.5550(84)$ & $0.437(15)$ & $0.671986(85)$ & $0.69449(33)$ \\
$2/5$ & $1.5250(86)$ & $0.356(17)$ & $0.66057(15)$ & $0.68326(33)$ \\
$1/3$ & $1.5000(96)$ & $0.288(17)$ & $0.65218(17)$ & $0.67443(27)$ \\
$1/4$ & $1.490(15)$ & $0.2056(75)$ & $0.64999(20)$ & $0.67088(73)$ \\
$1/5$ & $1.4600(90)$ & $0.1653(69)$ & $0.64151(17)$ & $0.66374(20)$ \\
$0.18$ & $1.4500(91)$ & $0.1513(50)$ & $0.639007(46)$ & $0.66121(31)$ \\

    \hline
    \end{tabular}
    \caption{Results for $\beta_\mathrm{iso}=1.7$ with $r_0/a_s$ matched to $r_\mathrm{iso}=3.462(40)$. The data are taken from the N0ET analysis chain.
    }
    \label{tab:res1.7}
\end{table}

\begin{table}[]
    \centering
    \begin{tabular}{lrrrr}
    \hline
    $\xi_\mathrm{input}$ & $\beta_\mathrm{match}$ & $\xiren$ & $P_\mathrm{match}(L=16)$ & $P_\mathrm{match}(L=3)$\\
\hline$4/5$ & $1.600(10)$ & $0.770(55)$ & $0.70256(17)$ & $0.72695(39)$ \\
$2/3$ & $1.565(11)$ & $0.614(32)$ & $0.68209(18)$ & $0.70697(41)$ \\
$1/2$ & $1.490(11)$ & $0.442(28)$ & $0.64864(14)$ & $0.67443(36)$ \\
$2/5$ & $1.480(14)$ & $0.340(17)$ & $0.644819(91)$ & $0.66879(33)$ \\
$1/3$ & $1.450(10)$ & $0.280(13)$ & $0.63424(11)$ & $0.65885(29)$ \\
$1/4$ & $1.420(14)$ & $0.2072(86)$ & $0.62530(16)$ & $0.64942(69)$ \\
$1/5$ & $1.420(10)$ & $0.1610(71)$ & $0.62759(21)$ & $0.65129(20)$ \\
$0.18$ & $1.405(10)$ & $0.1474(36)$ & $0.622708(27)$ & $0.64660(40)$ \\

\hline
    \end{tabular}
    \caption{Results for $\beta_\mathrm{iso}=1.65$ with $r_0/a_s$ matched to $r_\mathrm{iso}=3.231(52)$.
    The data are taken from the N0ET analysis chain.
    }
    \label{tab:res1.65}
\end{table}

\begin{figure}
    \centering
    \includegraphics[width=\linewidth]{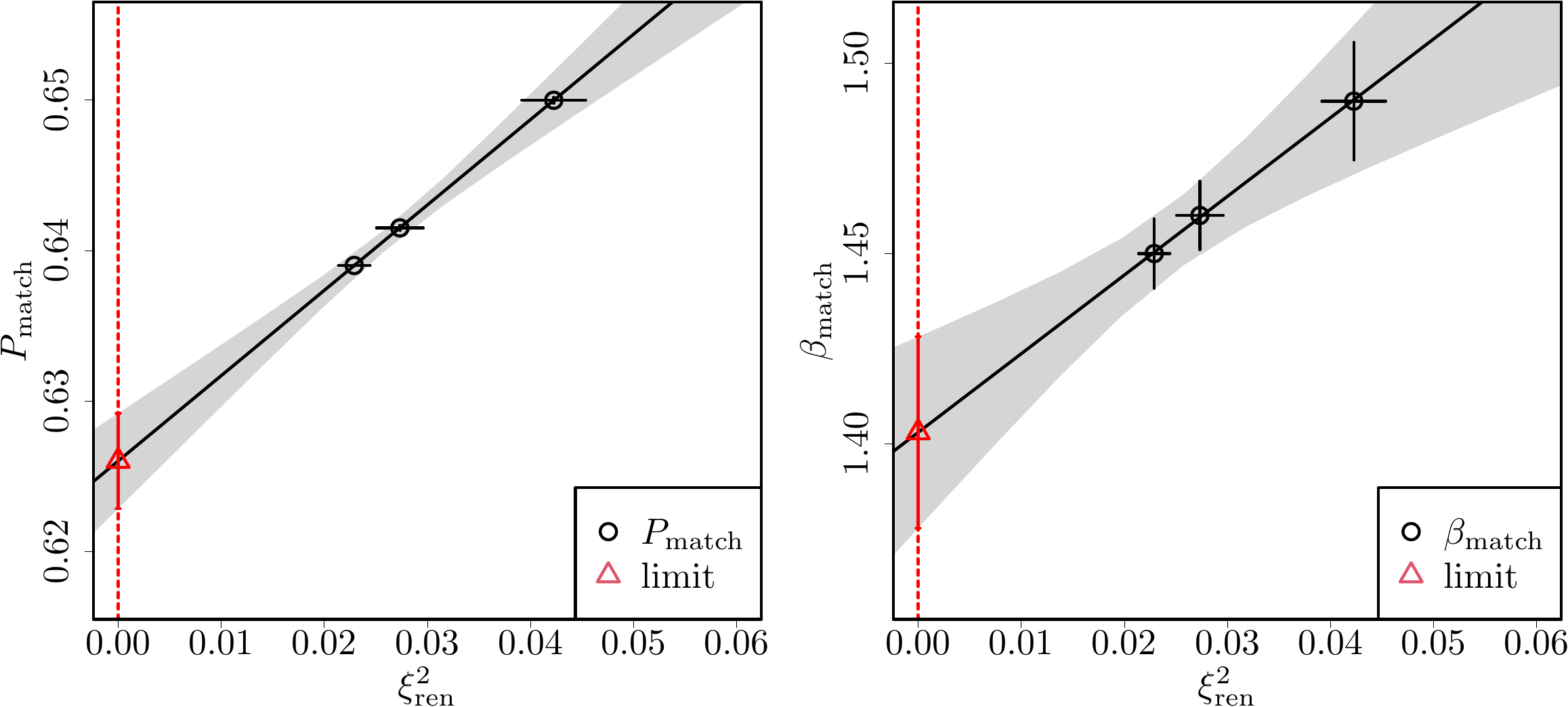}
    \caption{
The temporal continuum limit of $P_\mathrm{match}$ and $\beta_\mathrm{match}$. The points are the result of analysis chain N0ET. The plaquette is measured at $L=16$.
The fit is a linear fit including all three points corresponding to $\xi_\mathrm{input}=(0.18, 1/5, 1/4)$ and $\beta_\mathrm{iso}=1.7$.
}
    \label{fig:contlimitpbeta}
\end{figure}

\begin{figure}
    \centering
    \includegraphics[width=\linewidth]{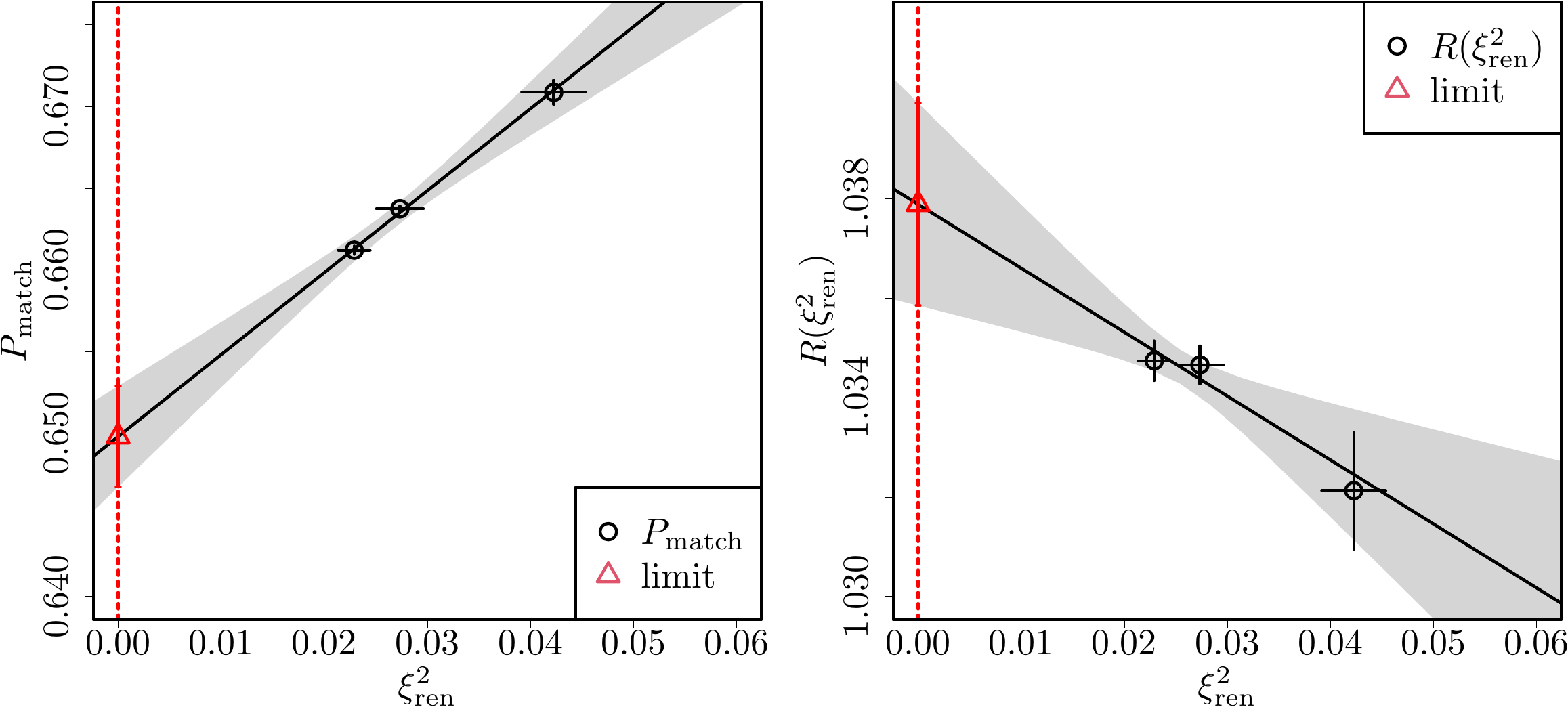}
    \caption{
The temporal continuum limit of $P_\mathrm{match}$ and $R(\xiren^2$). The points are the result of analysis chain N0ET. The plaquette is measured at $L=3$.
The fit is a linear fit including all three points. $\xi_\mathrm{input}=(0.18, 1/5, 1/4)$ and $\beta_\mathrm{iso}=1.7$.
}
    \label{fig:contlimitp3ratio}
\end{figure}

The next step in our analysis is the continuum limit in time direction, for which we show examples in \cref{fig:contlimitpbeta} for $\beta_\mathrm{match}$ and $P_\mathrm{match}$ with $L=16$ and $\beta_\mathrm{iso}=1.7$ as functions of $\xiren^2$, corresponding to
an extrapolation in the cA-ET set
with $\xiinput\in\{0.18, 1/5, 1/4\}$ and $n_p=1$ (see \cref{tab:groupcontlimit}, i.e.\ extrapolations linear in $\xiren^2$).
The analogous extrapolations with $L=3$ for $P$ and the ratio $R$ \cref{eq:ratio} are shown in \cref{fig:contlimitp3ratio}.
In both figures, the best fits are represented by the solid lines and its statistical uncertainties by the shaded bands. The red point marks the continuum limit.

\begin{figure}
    \centering
    \includegraphics[width=\linewidth]{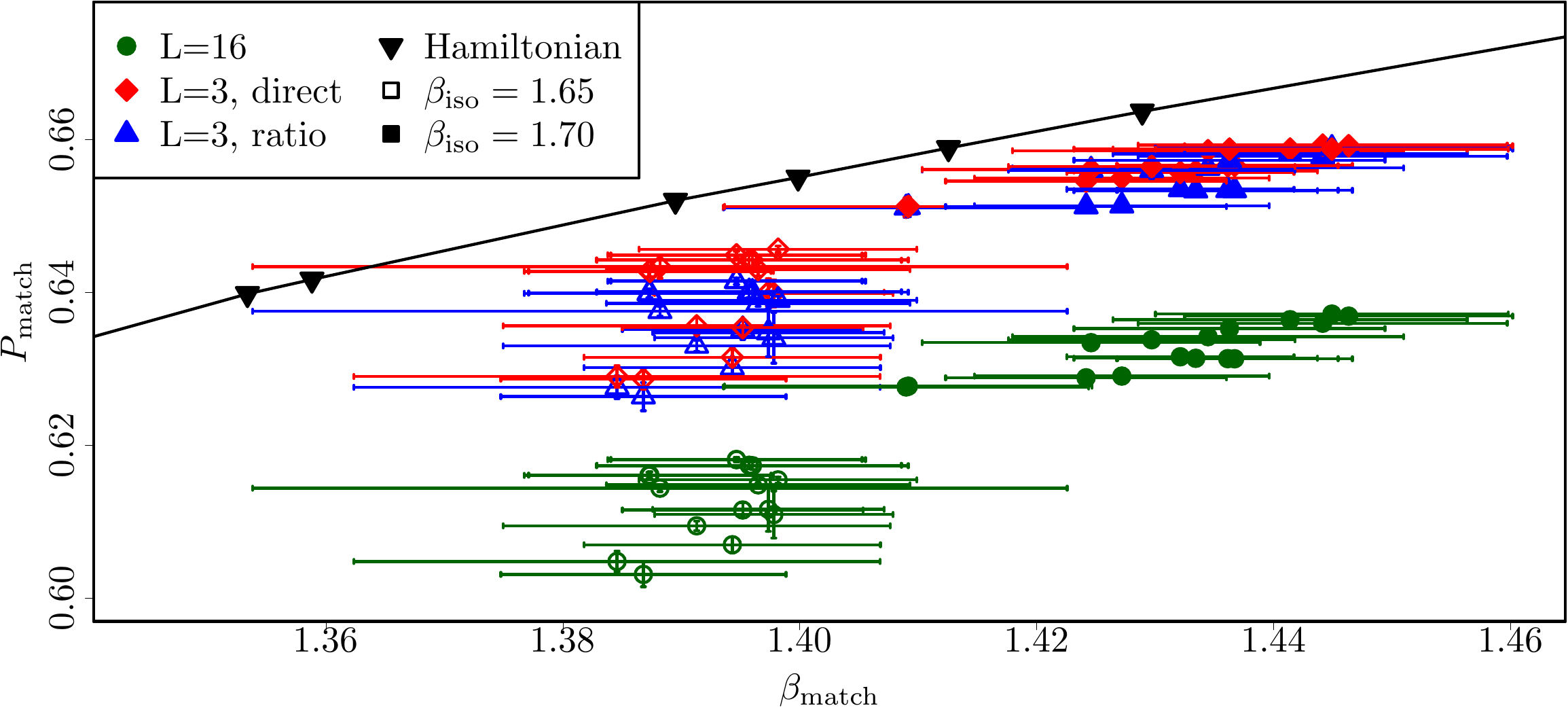}
    \caption{All continuum limits in the set cA-ET.
    The error bars show the combination of the statistical error and the error due to choosing the plateau boundaries.
    The open (filled) symbols correspond to $\beta_\text{iso}=1.65$ $(1.70)$.
    We display several different continuum limits:
    The limit at $L=16$, the direct fit of the $L=3$ data, and the result obtained for $L=3$ by using the ratio $R=P(L=3)/P(L=16)$.
    The Hamiltonian results are a combination of truncations $l=3$ and $l=4$, interpolated linearly.
    }
    \label{fig:allcombos}
\end{figure}

Taking into account now all analysis chains and the different sets of extrapolations leaves us with four sets of pairs of plaquette- and $\beta$-values in the temporal continuum limit. These pairs are visualised in \cref{fig:allcombos} for the set cA-ET.
The results of the Hamiltonian simulations are represented by the (black) upside-down triangles, connected by solid (black) lines to guide the eye.
The continuum limits for $L=16$ are represented by the green circles.
The results of the extrapolation of the plaquette measured at $L=3$ are shown as red diamonds, and the results obtained by extrapolating $R(\xiren^2)$ and multiplying with the $L=16$ result are shown as blue triangles.
The open (filled) symbols correspond to $\beta_\text{iso}=1.65$ $(1.70)$.
The error bars show the combination of the statistical error and the systematic error from the calculation of the potential.
First, we observe that the $L=16$ results without finite size corrections do not match the Hamiltonian results.
However, once $L=3$ is considered the MC results are much closer to the Hamiltonian results.

\begin{figure}
    \centering
    \includegraphics[width=\linewidth]{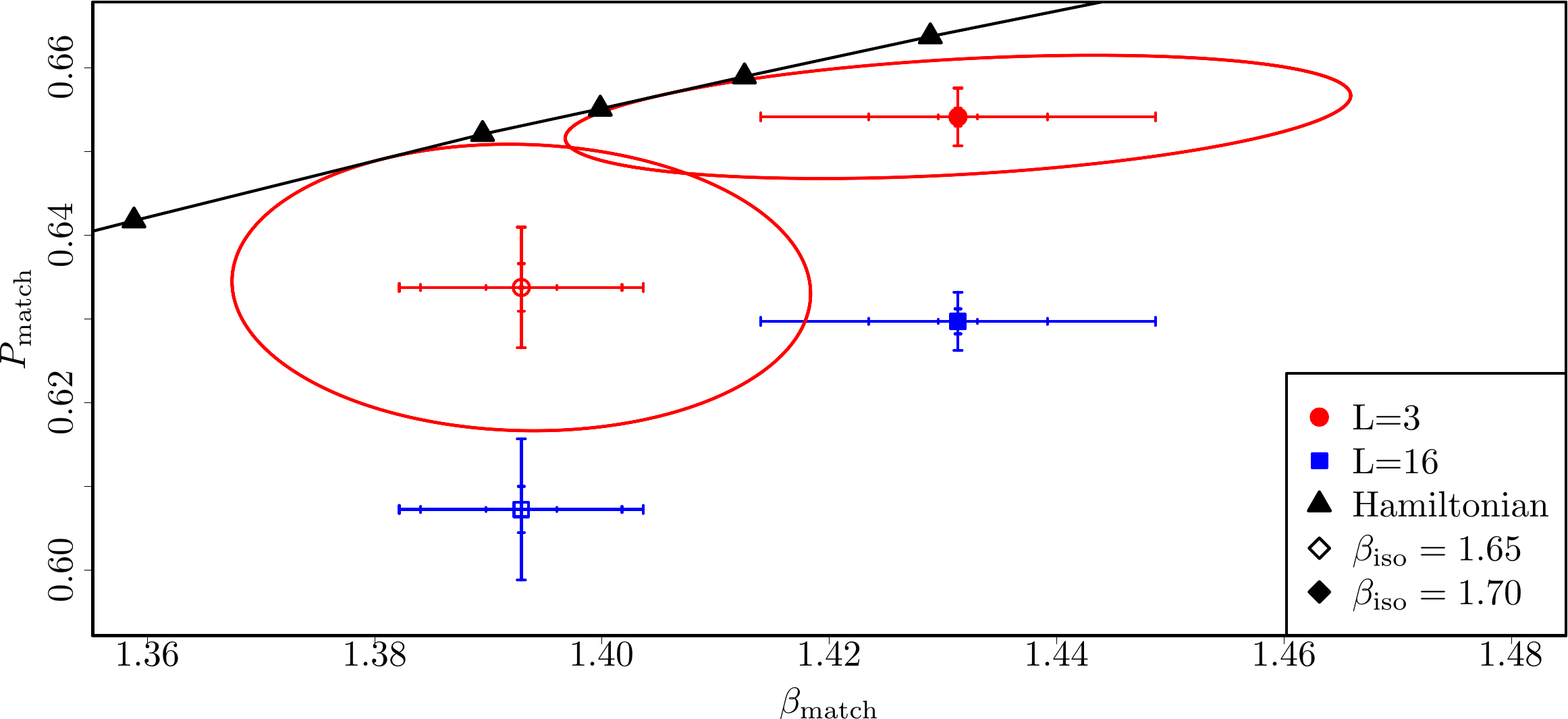}
    \caption{Combined continuum limits for the plaquette and $\beta$.
    The error markers correspond to the statistical error, the systematic error from choosing the potential plateaus, and the systematic error of the spread of the different continuum limit fits.
    In some cases the statistical error is so small it is not visible.
    The results are from the combination of all the fits in the set cA, with $P(L=3)$ calculated directly.
    The open (filled) symbols correspond to $\beta_\text{iso}=1.65$ $(1.70)$.
    The ellipse is determined with the procedure explained in \cref{subsec:touchellipse}.
    Further explanations are in the text.
    }
    \label{fig:finalwithellipse}
\end{figure}

Finally, we combine the continuum limits for all different sets of parameters and analysis chains as discussed in \cref{sec:temp-contlimit}.
The result of analysis set cA and the direct extrapolation of the small volume result is shown in \cref{fig:finalwithellipse}.
The black triangles are the combined $l=3$ and $l=4$ Hamiltonian results, joined by a linear interpolation.
The blue squares represent the Lagrangian results at $L=16$ and the red circles the results at $L=3$.
The open (filled) symbols correspond to $\beta_\text{iso}=1.65$ $(1.70)$.
The left (open) points correspond to $\beta_\text{iso}=1.65$ and show a residual $1.49\sigma$ deviation from the Hamiltonian curve. For the right (filled) points with $\beta_\text{iso}=1.70$ the deviation amounts to $1.88\sigma$.

The ellipse in \cref{fig:finalwithellipse} indicates a lower correlation between $\beta$ and $P$ than one might expect from the spread of the points in \cref{fig:allcombos}. This is because the points in \cref{fig:allcombos} correspond to fits with the same parameters for $P$ and $\beta$, but in the final result, also the correlations between fits with different parameters enter, and these are lower than the correlations between the fits with the same parameters.
The confidence ellipse of the Lagrangian result was calculated as described in \cref{sec:compareLandH,sec:confidencelevel}.

\begin{figure}
    \centering
    \includegraphics[width=\linewidth]{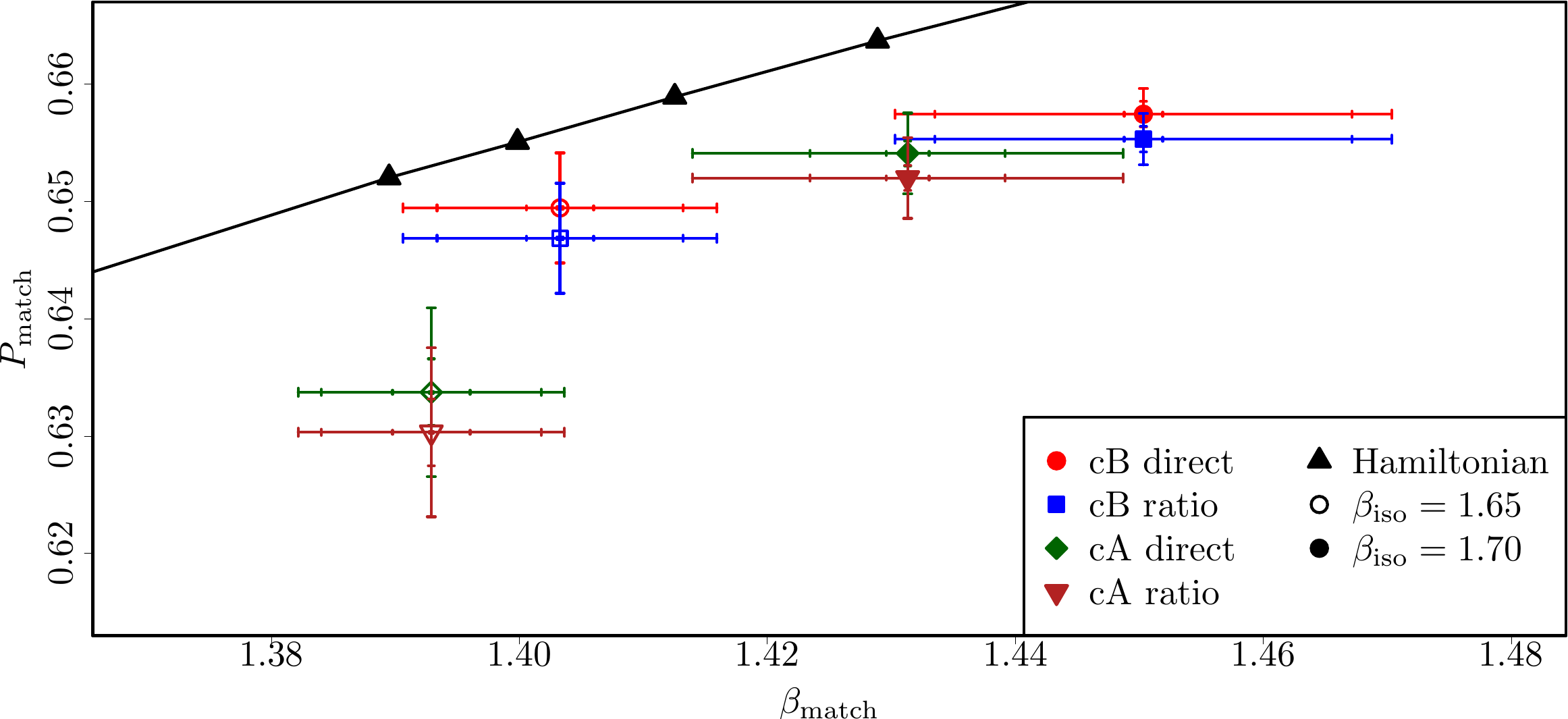}
    \caption{Combined continuum limits for the plaquette and $\beta$ for every combination of extrapolation set and small volume extrapolation.
    Further explanations are in the text.
    }
    \label{fig:finalallgroups}
\end{figure}

A comparison of all analysis sets and of the two ways of implementing the small volume limit are shown in \cref{fig:finalallgroups}.
The black triangles represent again the combined $l=3$ and $l=4$ Hamiltonian results, joined by a linear interpolation.
The red circles are the results of the cB extrapolation set with the direct extrapolation at $L=3$, the blue squares are the same set but with the small volume effects determined by the ratio $R$.
The green diamonds correspond to the cA direct extrapolation and the maroon upside-down triangles to the ratio extrapolation of the same extrapolation set.
The open (filled) symbols correspond to $\beta_\text{iso}=1.65$ $(1.70)$.

The two ways of calculating the small volume effects are fully compatible.
The results of the two extrapolation sets lead to compatible results for $\betamatch$, the results for $P_\text{match}$ are compatible for $\beta_\text{iso}=1.7$ and deviate by less than $2\sigma$ for $\beta_\text{iso}=1.65$.

The results for the matching $\beta$ are given in \cref{tab:finalcontlimbeta}, the plaquettes at the matching point are given in \cref{tab:finalresultsallchainscA,tab:finalresultsallchainscB} for the extrapolation sets cA and cB respectively.
The four errors we quote correspond to statistical only, the systematic error from calculating the potential energies, the systematic error from the spread of the different continuum limit results, and the total error, added in quadrature.

\begin{table}[htbp]
    \centering
    \begin{tabular}{lll}
    \hline
    $\beta_\text{iso}$ & set & $\betamatch$ \\
    \hline
$1.65$ & cB & $1.403(06)(10)(06)[13]$ \\
$1.7$  & cB & $1.450(06)(17)(08)[20]$ \\
$1.65$ & cA & $1.393(06)(08)(04)[11]$ \\
$1.7$  & cA & $1.431(05)(10)(13)[17]$ \\
\hline
    \end{tabular}
    \caption{The coupling constant $\beta$ at the matching point. For explanation of the errors given see in the text.}
    \label{tab:finalcontlimbeta}
\end{table}

\begin{table}[htbp]
    \centering

\begin{tabular}{l|llll}
\hline
$\beta_\text{iso}$ & $1.65$ & $1.7$\\
\hline
set & cA & cA\\
$P(L=16)$ & $0.6073(06)(48)(69)[84]$ & $0.6297(14)(18)(26)[35]$\\
$P(L=3)$ direct & $0.6337(07)(45)(56)[72]$ & $0.6541(10)(16)(29)[34]$\\
$\dev$ dir. & $1.49$ & $1.88$\\
$P(L=3)$ ratio & $0.6303(07)(50)(72)[88]$ & $0.6520(15)(18)(28)[36]$\\
$\dev$ rat. & $1.87$ & $1.87$\\
\hline

    \end{tabular}
    \caption{The different plaquette observables at the matching point for the extrapolation set cA. For explanation of the errors given see in the text.
    $\dev$ measures the difference between the Lagrangian and Hamiltonian simulations as explained in \cref{subsec:conflevelfromradius}.
    }
    \label{tab:finalresultsallchainscA}
\end{table}

\begin{table}[htbp]
    \centering

\begin{tabular}{l|llll}
\hline
$\beta_\text{iso}$ & $1.65$ & $1.7$\\
\hline
set & cB & cB\\
$P(L=16)$ & $0.6233(04)(11)(67)[67]$ & $0.6334(07)(18)(27)[34]$\\
$P(L=3)$ direct & $0.6495(03)(07)(46)[47]$ & $0.6574(07)(14)(15)[22]$\\
$\dev$ dir. & $1.57$ & $0.61$\\
$P(L=3)$ ratio & $0.6469(04)(12)(69)[70]$ & $0.6553(07)(19)(28)[34]$\\
$\dev$ rat. & $0.66$ & $1.77$\\
\hline

    \end{tabular}
    \caption{Same as \cref{tab:finalresultsallchainscA}, but for extrapolation set cB
    }
    \label{tab:finalresultsallchainscB}
\end{table}

\section{Discussion}
\label{sec:discussion}

In \cref{fig:thermalizationtriple} we see that the number of sweeps it takes to reach an equilibrium state grows with decreasing $\xiinput$, eventually leading to critical slowing down.
This is also seen in the autocorrelation times in \cref{tab:usedconfigsL16,tab:usedconfigsL3} and \cref{fig:autocorrelation}, they grow with decreasing anisotropy.
For $\xiinput \geq 1/4$, we were able to simulate with the standard Metropolis algorithm, whereas for even smaller anisotropies, the autocorrelation was too large and we were unable to even achieve thermalisation.
The combination of heatbath and overrelaxation algorithms mitigated the critical slowing down enough to make simulations at $\xiinput=1/5, 0.18$ feasible.
However, for $\xiinput<0.18$ we could not reach equilibrium in reasonable simulation times.

This restricts of course how close we could get with our simulations to the temporal continuum limit.
Still, we are confident that our procedure of following different analysis chains, and of using different sets of temporal continuum extrapolations leads to reliable estimates of the uncertainties.
In fact, the total uncertainty is in almost all results dominated by systematic uncertainties.
Moreover, statistical fluctuations are more likely to average out in our procedure.

It is also reassuring that the extrapolation sets cA and cB lead to compatible results: we recall that the most significant difference between the two is the in- or exclusion of the data at the smallest $\xiinput$-value.
Therefore, the data at the smallest $\xiinput$-value confirm our temporal continuum limit results, but are not strictly necessary.

Our results in the temporal continuum limit are in agreement with the Hamiltonian results within two $\sigma$; the largest deviation is $1.87 \sigma$.
This indicates in general that taking the temporal continuum limit in the Euclidean $(2+1)$ dimensional lattice theory is equivalent to the Hamiltonian lattice theory for the specific lattice action and Hamiltonian quoted in the introduction.

Still, there might be a systematic effect unaccounted for, because all our extrapolation results lie below the Hamiltonian curve.
One possible explanation for this systematic deviation could be the truncation on the Hamiltonian side.
This seems not to be the case because larger $l$-values tend to push the plaquette values up at fixed $\beta$.
Despite the discussion from above, we certainly cannot be 100\% sure that we are close enough to the temporal continuum limit, which might offer one explanation for the systematic deviation.
However, this will need further investigation in the future.

\section{Summary and Outlook}
\label{sec:outlook}

We have performed the temporal continuum limit in a U$(1)$ lattice gauge theory using stochastic simulations in the Lagrangian formalism. We performed this temporal continuum limit using the anisotropic lattice formulation starting with two $\beta$-values from the isotropic side.
Trajectories of constant spatial lattice spacing are defined by keeping the Sommer parameter $r_0/a_s$ fixed.
The so obtained temporal continuum results for the plaquette and the coupling $\beta$ are compared to results from a direct Hamiltonian simulation.
We find general agreement within two $\sigma$ between Hamiltonian and extrapolated Lagrangian results.
As discussed in the previous section, the deviation is systematic towards lower plaquette values for the extrapolated results, for which we currently do not have a good explanation.
\comment{Our procedure allows us to calculate observables at the same lattice spacing $a_s$ in both theories.
For the comparison between the theories, we are not interested in the limit $a_s \to 0$, but instead we use the matching point to combine advantages of both theories at fixed lattice spacing.}

There are several immediate extensions that we leave for the future:
On the path integral side, it is possible to use other parameters to set the scale, e.g.\ the time $\tau_0$ from the gradient flow, or a fermion mass or decay constant in a fermionic theory.
In a fermionic theory we could also use other matching variables, like the mass gap.

On the Hamiltonian side, the goal is to simulate larger lattices, which simplifies the matching. This can be achieved by other methods beyond exact diagonalization,  such as Tensor Networks and future Quantum Computers.
Larger lattices in future simulations would make other matching variables beyond the plaquette possible, and would reduce the need for finite volume extrapolations.

An extension to higher dimensions or other lattice gauge theories, in particular QCD, is conceptually straightforward, but will be computationally demanding on both the Lagrangian and the Hamiltonian side.

\section*{Acknowledgements}
The work on this project was supported by the Deutsche Forschungsgemeinschaft (DFG, German Research Foundation) as part of the CRC 1639 NuMeriQS – project no. 511713970 and as part of NRW-FAIR by the MKW NRW under the funding code NW21-024-A.
This work is supported with funds from the Ministry of Science, Research and Culture of the State of Brandenburg within the Center for Quantum Technology and Applications (CQTA).
\begin{center}
    \includegraphics[width = 0.05\textwidth]{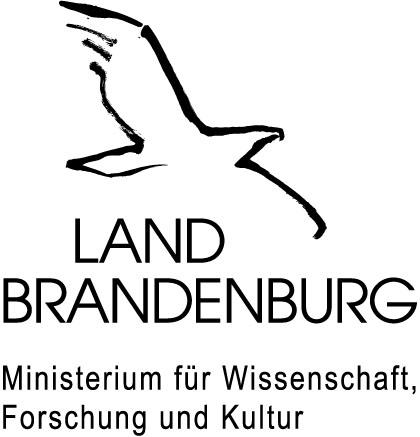}
\end{center}

This work is supported by the European Union’s Horizon Europe Framework Programme (HORIZON) under the ERA Chair scheme with grant agreement no. 101087126.

The authors gratefully acknowledge the granted access to the Marvin cluster hosted by the University of Bonn.
We thank A.~Crippa for helpful discussions and for providing additional results of the Hamiltonian simulations. We also thank M.~Garofalo for helpful discussions.

This version of the article has been accepted for publication, after peer review but is
not the Version of Record and does not reflect post-acceptance improvements, or any corrections. The Version of Record is available online at: \href{http://dx.doi.org/10.1140/epjc/s10052-025-14923-2}{http://dx.doi.org/10.1140/epjc/s10052-025-14923-2}.

\appendix

\section{Autocorrelation}
\label{sec:autocorrelation}

A common way to deal with autocorrelation is to block the data before bootstrapping it.
However, our results showed that this lead to a bad estimation of the covariance matrix, and thus unreliable results of the potential energies.

We take care of the autocorrelation in a different way:
The potential energies are still determined with bootstrapping, but with block length 1.
At the same time, we determine the autocorrelation of the original data with the UWerr-algorithm \cite{Wolff:2003sm} implemented in~\cite{kostrzewa_2024_14800686}.
We then rescale the results of the bootstrapping to take the autocorrelation into account.
The error is rescaled with $2\tau_\text{int}$ and each bootstrap sample $x$ is rescaled to be $x_\mathrm{new} = x_\mathrm{old} + (x_\mathrm{old} - \mu) \cdot (2 \cdot \tau_\text{int} - 1)$, with $\mu$ the unbiased mean.
This ensures that the difference of the bootstrap sample to the unbiased mean is $2\tau_\text{int}$ times as large, so the error increases by the required amount.

We also draw bootstrap samples of the plaquette.
There, we set the block length to $4\tau_\text{int}^2$ to take the autocorrelation into account.

\section{Using AIC to determine the potential points}
\label{sec:aic}

The potential points are computed with the help of the Akaike Information Criterion (AIC)~\cite{Borsanyi:2020mff}.

We fit the ratios from \cref{eq:normalpotential,eq:sidewayspotential} to a constant in the region $(t_1, t_2)$.
All possible combinations with $t_1 > 1$ and $t_2-t_1 > 2$ are used, and each fit result is assigned the weight
\begin{equation}
w = \exp \left( -\frac {1}{2} \cdot \left ( \chi^2 + 2 - (t_2 - t_1) \right) \right)
\label{eq:weight-aic}
\end{equation}
with $\chi^2$ the sum of residues of the fit.

The weights are normalized to one.
We assume the results are normally distributed and combine the weights, means $\mu$ and standard deviations $\sigma$ to give the cumulative distribution function (cdf)

\begin{equation}
\text{cdf}(y) = \sum_i w_i \cdot \frac{1}{2} \left( 1 + \frac{\mathrm{erf}(y-\mu_i)}{\sqrt{2 \sigma_i^2}}\right)
\end{equation}
where $\mathrm{erf}$ is the error function and $y$ is the potential energy.

The cdf is used to find the median $q_{50}$, the $16\%$ quantile $q_{16}$ and the $84\%$ quantile $q_{84}$ of the distribution of the masses.
The procedure is repeated for each bootstrap sample.

The medians of the bootstrap samples represent the statistical error.
However, $q_{16}$ and $q_{84}$ also give us information about the total error, including the systematic uncertainty in choosing the correct boundaries of the effective mass fit.

We use $\sigma_\mathrm{comb}=\frac{1}{2}(q_{84}-q_{16})$ of the result of the original data as an estimate of the total error, and $\sigma_\mathrm{stat} = \mathrm{sd} (q_{50, \text{boot}})$ as an estimate of the statistical error.

To use the total error in the further analysis, we rescale each bootstrap sample so that it is $\frac{\sigma_\mathrm{comb}}{\sigma_\mathrm{stat}}$ further away from the mean, similar to the rescaling to take into account the autocorrelation as described in \cref{sec:autocorrelation}.
We can extract the systematic error from $\sigma_\mathrm{comb}^2 = \sigma_\mathrm{stat}^2 + \sigma_\mathrm{pot}^2$.

\section{Determining confidence level}
\label{sec:confidencelevel}

\subsection{Getting a touching ellipse}
\label{subsec:touchellipse}

We want to determine the difference between the Hamiltonian and Lagrangian results and do this geometrically with an ellipse.

We piece wise linearly interpolate the Hamiltonian results and determine the equation for each piece.

We know the Lagrangian result $(\beta, P)$ and its errors $(\sigma_\beta, \sigma_P)$. This is the centre point of the ellipse, and the ratio of the errors is the ratio of the major axes of the ellipse.
The angle $\phi$ of the ellipsis is given by $\tan (2\phi) = \frac{2\rho \sigma_\beta  \sigma_P}{\sigma_\beta^2 - \sigma_P^2}$, with $\rho$ the correlation coefficient of the bootstrap samples of $\beta$ and $P$~\cite{CowanGlen1998SDA}.

The ellipse can be written in the general form as
\begin{equation}
A_{xx}x^2+2A_{xy}xy+A_{yy}y^2 + 2B_x x + 2B_y y +C = 0
\label{eq:ellipse}
\end{equation}

or, written in matrix form

\begin{equation}
\begin{pmatrix} x \\ y \\ 1 \end{pmatrix}^T
\begin{pmatrix}
    A_{xx} & A_{xy} & B_x \\
    A_{xy} & A_{yy} & B_y \\
    B_x & B_y & C
\end{pmatrix}
\begin{pmatrix} x \\ y \\ 1 \end{pmatrix}
= 0 = \Tilde{X}^T A \Tilde{X}
\label{eq:ellipsematrix}
\end{equation}

with

\begin{align}
A_{xx} &= \sigma_\beta^2 \cdot \sin^2(\phi) + \sigma_P^2 \cdot \cos^2(\phi) \nonumber \\
A_{xy} &= (\sigma_P^2 -  \sigma_\beta^2) \cdot \sin(\phi)  \cdot \cos(\phi) \nonumber \\
A_{yy} &= \sigma_\beta^2 \cdot \cos^2(\phi) + \sigma_P^2 \cdot \sin^2(\phi) \nonumber \\
B_x &= -A_{xx} \cdot \beta - A_{xy} \cdot P \nonumber \\
B_y &= -A_{xy} \cdot \beta - A_{yy} \cdot P \nonumber \\
C &= A_{xx} \cdot \beta^2 + 2A_{xy} \beta P + A_{yy} P^2 - \sigma_\beta^2 \sigma_P^2\, .
\label{ellipsematrixelements}
\end{align}

Here we have set the major axes equal to the errors, as  is the starting case for our calculation.

We interpret the Hamiltonian interpolation as a polar to the ellipse.
The interpolation can be expressed as a line of the form $Dx+Ey+F=0=B^T\Tilde{X}$.
This curve is invariant under a rescaling of $B$.

A polar to a point $P$ can be written as $P^TA\Tilde{X}=0$~\cite{Fortuna:2016}.

Setting the two descriptions of the polar equal to each other, we get
\begin{equation}
    P^TA=B^T \Leftrightarrow AP=B \Leftrightarrow P=A^{-1}B\, .
    \label{eq:polar}
\end{equation}

With
\begin{equation}
\begin{pmatrix}
    x_P \\ y_P \\ z_P
\end{pmatrix}
=
\begin{pmatrix}
    A_{xx} & A_{xy} & B_x \\
    A_{xy} & A_{yy} & B_y \\
    B_x & B_y & C
\end{pmatrix}^{-1}
\begin{pmatrix}
    D \\ E \\ F
\end{pmatrix}
\label{eq:determinepolar}
\end{equation}
the pole has the coordinates $(\frac{x_P}{z_P}, \frac{y_P}{z_P})$, where we chose $z_P$ as our rescaling factor for the polar equation.
If the polar is a tangent to the ellipse, the pole is the touching point and directly on the ellipse~\cite{Fortuna:2016}.
In that case, it fulfils the ellipse equation \cref{eq:ellipse}.

For determining the matching level, we keep the angle, centre point and ratio of the major axes fixed, but vary the length of the major axes by setting them to $r \sigma_P$ and $r \sigma_\beta$.

Finding the matching level and radius is reduced to a root-finding procedure:
We keep everything except the radius $r$ fixed, and vary the radius until the pole is on the ellipse, and thus the polar is a tangent at the radius $r^*$.
We call the ellipse corresponding to the radius $r^*$ the matching ellipse, because at this radius it touches the interpolation of the Hamiltonian results.

We do this for every interpolated piece, and then select only the piece(s) for which the matching point is on the piece itself.
This usually only yields one matching point and level of deviation, but in case several points match, we select the one with the lower $r^*$.

\subsection{Getting a confidence level from the radius}
\label{subsec:conflevelfromradius}

A non-tilted ellipse centred at the origin can be written as

\begin{equation}
\left(\frac{x}{\sigma_x}\right)^2
+
\left(\frac{y}{\sigma_y}\right)^2 = s\,\,.
\label{eq:nontiltedellipse}
\end{equation}

The probability distribution for $x, y$ is
\begin{equation}
P(x,y) = \frac{dx dy }{(2\pi) \sigma_x \sigma_y}
e^{-\frac{x^2}{2\sigma_x^2} -\frac{y^2}{2\sigma_y^2}}\,\,.
\label{eq:pdfellipse}
\end{equation}

Integrating over this in the area of the ellipse gives us the probability $p$ that a point is inside the ellipse $p=1-\exp\left(-0.5\cdot s\right)$.

We can apply this to a tilted ellipse as well, and identify $\beta, P$ with $x, y$ and $r^{*2}$ with $s$.

From the matching ellipse, we can thus calculate the probability $p$ that a point from the Lagrangian distribution is inside the ellipse and not a match to the Hamiltonian.

We convert this probability into units of the standard deviation of the normal distribution.

We determine the deviation level, $\dev$, by setting $p=\Phi(\dev)-\Phi(-\dev)$, with $\Phi$ the cumulative density function of the normal distribution.
We determine $\dev$ by a root-finding procedure on

\begin{equation}
\left(\Phi(\dev)-\Phi(-\dev)\right) - \left(1-\exp\left(-0.5\cdot r^{*2}\right)\right) = 0\,\,.
\label{eq:definitiondev}
\end{equation}

\section{Systematics of continuum limit}

\begin{figure}[htbp]
     \centering
     \begin{tikzpicture}[node distance=2cm]

\node (ensemble) [rectangle, 
minimum width=2cm, 
minimum height=1cm, 
text centered, 
text width=2cm, 
draw=black, 
fill=orange!30] {Ensembles};

\node (resES) [rectangle, 
minimum width=1cm, 
minimum height=1cm, 
text centered, 
text width=1.5cm, 
draw=black, 
fill=orange!30, 
below of=ensemble, 
xshift=-1.5cm] {ES};

\node (resET) [rectangle, 
minimum width=1cm, 
minimum height=1cm, 
text centered, 
text width=1.5cm, 
draw=black, 
fill=orange!30, 
below of=ensemble, 
xshift=+1.5cm] {ET};
\draw [very thick, dashed, color=blue] (-0.5,-0.5) -- (-1.5, -1.5);
\draw [very thick, dashed, color=blue] (-0.7,-0.5) -- (-1.7, -1.5);
\draw [very thick, dotted, color=blue] (+0.5,-0.5) -- (+1.5, -1.5);
\draw [very thick, dotted, color=blue] (+0.7,-0.5) -- (+1.7, -1.5);

\draw [very thick, dashed, color=blue] (-1.0,-2.5) -- (-1.0, -3.5);
\draw [very thick, dashed, color=blue] (-2.0,-2.5) -- (-2.0, -3.5);
\draw [very thick, dotted, color=blue] (+1.0,-2.5) -- (+1.0, -3.5);
\draw [very thick, dotted, color=blue] (+2.0,-2.5) -- (+2.0, -3.5);

\draw [very thick, dashed, color=red] (-1.0, -3.5) -- (-1.25, -5);
\draw [very thick, dashed, color=black] (-1.0, -3.5) -- node[anchor=west] {cA} (-0.75, -5);
\draw [very thick, dashed, color=red] (-2.0, -3.5) -- node[anchor=east] {cB} (-2.25, -5);
\draw [very thick, dashed, color=black] (-2.0, -3.5) -- (-1.75, -5);
\draw [very thick, dashed, color=red] (-1.0, -3.5) -- (-1.1, -5);
\draw [very thick, dashed, color=black] (-1.0, -3.5) -- (-0.9, -5);
\draw [very thick, dashed, color=red] (-2.0, -3.5) -- (-2.1, -5);
\draw [very thick, dashed, color=black] (-2.0, -3.5) -- (-1.9, -5);

\draw [very thick, dotted, color=black] (+1.0, -3.5) -- (+1.25, -5);
\draw [very thick, dotted, color=red] (+1.0, -3.5) -- node[anchor=east] {cB} (+0.75, -5);
\draw [very thick, dotted, color=black] (+2.0, -3.5) -- node[anchor=west] {cA} (+2.25, -5);
\draw [very thick, dotted, color=red] (+2.0, -3.5) -- (+1.75, -5);
\draw [very thick, dotted, color=black] (+1.0, -3.5) -- (+1.1, -5);
\draw [very thick, dotted, color=red] (+1.0, -3.5) -- (+0.9, -5);
\draw [very thick, dotted, color=black] (+2.0, -3.5) -- (+2.1, -5);
\draw [very thick, dotted, color=red] (+2.0, -3.5) -- (+1.9, -5);

\node (combine) [rectangle, 
minimum width=6.8cm, 
minimum height=1cm, 
text centered, 
text width=5cm, 
draw=black, 
fill=orange!30, 
below of=ensemble, 
yshift=-3.5cm] {cont. limit results};

\draw[decorate, decoration = {brace, raise=0.2cm}, very thick] (+3.5, -6) --  (-3.5, -6);

\node (EScB) [fill=orange!30, 
very thick, 
circle, 
text centered,
text=red, 
dashed,
draw=red, 
below of=combine, 
xshift=-3 cm] {cB-ES};

\node (EScA) [fill=orange!30, 
very thick, 
circle, 
text centered, 
text=black,
draw=black, dashed,
below of=combine, 
xshift=-1 cm] {cA-ES};

\node (ETcB) [fill=orange!30, 
very thick, 
circle, 
text centered, 
draw=red, dotted,
below of=combine, 
xshift=+1 cm, 
text=red] {cB-ET};

\node (ETcA) [fill=orange!30, 
very thick, 
circle, 
text centered, 
draw=black, dotted,
below of=combine, 
xshift=+3 cm, 
text=black] {cA-ET};

\node (detxi) [below of=ensemble, yshift=1.2cm, xshift=6.5cm, text width=4cm] {determine $\xi_\text{ren}, r_0/a_s$ in 
analysis chains};
\node (dettraj) [below of=ensemble, yshift=-0.7cm, xshift=6.5cm, text width=4cm] {determine trajectory of constant 
$a_s$};
\node (contlim) [below of=ensemble, yshift=-2.2cm, xshift=6.5cm, text width=4cm] {extrapolate to cont. limit};

\node (staterrcB) [below of=EScB, fill=orange!30, 
very thick, 
rectangle, 
text centered, 
draw=red,dashed, 
text=red] {$\sigma_\text{stat}$};

\draw [very thick, dashed, color=red] (EScB) -- (staterrcB);

\node (staterrcA) [below of=EScA, fill=orange!30, 
very thick, 
rectangle, 
text centered, 
draw=black, dashed, 
text=black] {$\sigma_\text{stat}$};

\draw [very thick, dashed, color=black] (EScA) -- (staterrcA);

\node (toterrcB) [below of=ETcB, fill=orange!30, 
very thick, 
rectangle, 
text centered, 
draw=red, dotted,
text width=1.5cm,  
text=red] {$\sigma_\text{comb}$ $\sigma_\text{spread,tot}$ };

\draw [very thick, dotted, color=red] (ETcB) -- (toterrcB);

\node (toterrcA) [below of=ETcA, fill=orange!30, 
very thick, 
rectangle, 
text centered, 
draw=black, dotted,
text width=1.5cm,  
text=black] {$\sigma_\text{comb}$ $\sigma_\text{spread,tot}$ };

\draw [very thick, dotted, color=black] (ETcA) -- (toterrcA);

\draw[decorate, decoration = {brace, raise=0.2cm}, very thick] (+3.5, -10) --  (-3.5, -10);

\node (finalerrcA) [below of=staterrcB, fill=orange!30, 
very thick, 
rectangle, 
text centered, 
draw=black, 
xshift=1cm,
minimum height=1cm, 
minimum width=1cm, 
text width=1.5cm,  
text=black] {cA \\ $\sigma_\text{stat}$ \\ $\sigma_\text{pot}$ \\ $\sigma_\text{spread}$};

\node (finalerrcB) [below of=toterrcB, fill=orange!30, 
very thick, 
rectangle, 
text centered, 
draw=red,
xshift=1cm,
minimum height=1cm, 
minimum width=1cm, 
text width=1.5cm,  
text=black] {cB \\ $\sigma_\text{stat}$ \\ $\sigma_\text{pot}$ \\ $\sigma_\text{spread}$};

\node (collectsets) [below of=ensemble, yshift=-5.5cm, xshift=6.5cm, text width=4cm] {collect in sets};
\node (extracterr) [below of=ensemble, yshift=-7.5cm, xshift=6.5cm, text width=4cm] {extract errors from average and 
spread};
\node (finalerr) [below of=ensemble, yshift=-9.5cm, xshift=6.5cm, text width=4cm] {calculate individual errors};

\end{tikzpicture}
     \caption{Visualisation of the different steps of the analysis procedure.
     The procedure is described in detail in \cref{sec:temp-contlimit}.
     We display only four analysis chains (see \cref{tab:analysischains}) and four kinds of fit to the continuum limit (see \cref{tab:groupcontlimit}).
     The analysis goes from top to bottom, and we denote the inclusion (exclusion) of a systematic error from the potential by dotted (dashed) lines, and the inclusion (exclusion) of the smallest anisotropy in the fits for the continuum limit by black (red) lines.
     }
     \label{fig:flowchartanalysis_extended}
 \end{figure}
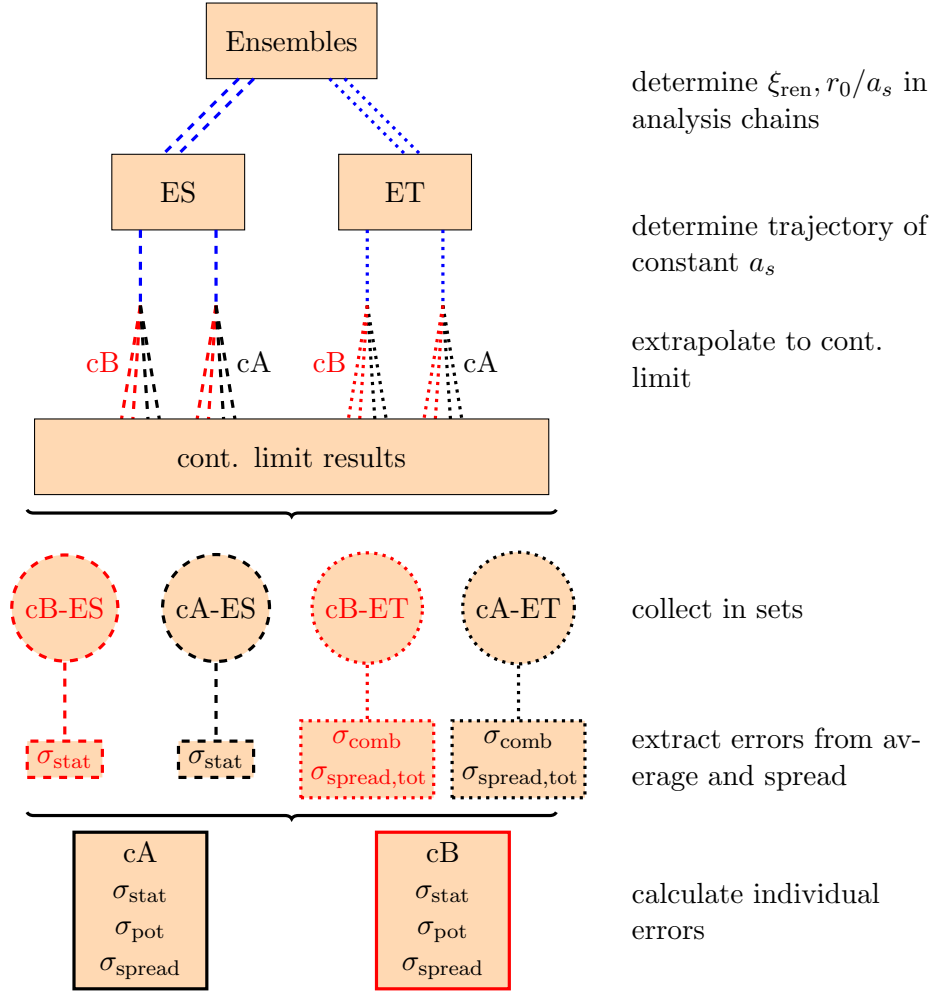

Our procedure to average over different ways of taking the continuum limit from  \cref{sec:temp-contlimit} is visualised in \cref{fig:flowchartanalysis_extended}.

\section{used configurations}
\label{sec:listconfigs}

The configurations are given in \cref{tab:usedconfigsL16} for $L=16$ and in \cref{tab:usedconfigsL3} for $L=3$.

\newpage

    \begin{longtable}{llllllll}
    \hline
    $\xiinput$ & $\beta$ & \# confs. & $L$ & $T$ & algo & \# sweeps & $\tau_\text{int}$\\
    \hline
$1$ & $1.65$ & $5500$ & $16$ & $16$ & met & $50$ & $0.490(13)$\\
$1$ & $1.7$ & $9748$ & $16$ & $16$ & met & $100$ & $0.497(10)$\\
$4/5$ & $1.59$ & $1500$ & $16$ & $20$ & met & $50$ & $0.516(65)$\\
$4/5$ & $1.6$ & $1500$ & $16$ & $20$ & met & $50$ & $0.585(95)$\\
$4/5$ & $1.615$ & $1500$ & $16$ & $20$ & met & $50$ & $0.499(45)$\\
$4/5$ & $1.63$ & $1500$ & $16$ & $20$ & met & $50$ & $0.489(44)$\\
$4/5$ & $1.64$ & $9749$ & $16$ & $20$ & met & $100$ & $0.490(10)$\\
$4/5$ & $1.65$ & $4749$ & $16$ & $20$ & met & $100$ & $0.502(39)$\\
$4/5$ & $1.6521$ & $1750$ & $16$ & $20$ & met & $50$ & $0.496(34)$\\
$4/5$ & $1.665$ & $1500$ & $16$ & $20$ & met & $50$ & $0.497(63)$\\
$4/5$ & $1.7$ & $4749$ & $16$ & $20$ & met & $100$ & $0.503(39)$\\
$4/5$ & $1.75$ & $4749$ & $16$ & $20$ & met & $100$ & $0.500(25)$\\
$2/3$ & $1.54$ & $28000$ & $16$ & $24$ & hb & $10$ & $0.581(26)$\\
$2/3$ & $1.555$ & $3750$ & $16$ & $24$ & met & $50$ & $0.500(16)$\\
$2/3$ & $1.56$ & $3750$ & $16$ & $24$ & met & $50$ & $0.500(16)$\\
$2/3$ & $1.565$ & $1500$ & $16$ & $24$ & met & $50$ & $0.519(66)$\\
$2/3$ & $1.58$ & $2749$ & $16$ & $24$ & met & $100$ & $0.550(63)$\\
$2/3$ & $1.595$ & $7749$ & $16$ & $24$ & met & $100$ & $0.533(36)$\\
$2/3$ & $1.6$ & $28000$ & $16$ & $24$ & hb & $10$ & $0.578(25)$\\
$2/3$ & $1.6075$ & $1750$ & $16$ & $24$ & met & $50$ & $0.487(62)$\\
$2/3$ & $1.7$ & $1500$ & $16$ & $24$ & met & $50$ & $0.491(57)$\\
$1/2$ & $1.49$ & $2749$ & $16$ & $32$ & met & $100$ & $0.600(76)$\\
$1/2$ & $1.515$ & $1500$ & $16$ & $32$ & met & $50$ & $0.513(75)$\\
$1/2$ & $1.525$ & $4750$ & $16$ & $32$ & met & $100$ & $0.610(61)$\\
$1/2$ & $1.53$ & $4749$ & $16$ & $32$ & met & $100$ & $0.562(51)$\\
$1/2$ & $1.54$ & $5750$ & $16$ & $32$ & met & $50$ & $0.569(52)$\\
$1/2$ & $1.55$ & $4749$ & $16$ & $32$ & met & $100$ & $0.516(40)$\\
$1/2$ & $1.555$ & $5750$ & $16$ & $32$ & met & $50$ & $0.563(47)$\\
$1/2$ & $1.56$ & $5750$ & $16$ & $32$ & met & $50$ & $0.567(49)$\\
$1/2$ & $1.575$ & $5750$ & $16$ & $32$ & met & $50$ & $0.523(41)$\\
$1/2$ & $1.7$ & $1500$ & $16$ & $32$ & met & $50$ & $0.62(11)$\\
$2/5$ & $1.45$ & $5500$ & $16$ & $40$ & met & $50$ & $0.840(98)$\\
$2/5$ & $1.46$ & $2500$ & $16$ & $40$ & met & $50$ & $0.80(13)$\\
$2/5$ & $1.47$ & $5500$ & $16$ & $40$ & met & $50$ & $0.727(78)$\\
$2/5$ & $1.48$ & $5500$ & $16$ & $40$ & met & $50$ & $0.678(70)$\\
$2/5$ & $1.495$ & $5500$ & $16$ & $40$ & met & $50$ & $1.28(18)$\\
$2/5$ & $1.511$ & $1750$ & $16$ & $40$ & met & $50$ & $0.66(11)$\\
$2/5$ & $1.52$ & $5500$ & $16$ & $40$ & met & $50$ & $0.802(91)$\\
$2/5$ & $1.525$ & $2500$ & $16$ & $40$ & met & $50$ & $0.74(11)$\\
$2/5$ & $1.55$ & $5500$ & $16$ & $40$ & met & $50$ & $0.729(78)$\\
$2/5$ & $1.57$ & $5500$ & $16$ & $40$ & met & $50$ & $1.17(16)$\\
$2/5$ & $1.7$ & $1500$ & $16$ & $40$ & met & $50$ & $1.81(52)$\\
$1/3$ & $1.43$ & $15000$ & $16$ & $48$ & met & $50$ & $1.32(13)$\\
$1/3$ & $1.44$ & $15000$ & $16$ & $48$ & met & $50$ & $1.23(11)$\\
$1/3$ & $1.45$ & $7000$ & $16$ & $48$ & met & $50$ & $1.19(15)$\\
$1/3$ & $1.46$ & $15000$ & $16$ & $48$ & met & $50$ & $1.118(97)$\\
$1/3$ & $1.47$ & $4750$ & $16$ & $48$ & met & $100$ & $0.782(93)$\\
$1/3$ & $1.48$ & $2749$ & $16$ & $48$ & met & $100$ & $0.93(15)$\\
$1/3$ & $1.4814$ & $3500$ & $16$ & $48$ & met & $50$ & $1.38(25)$\\
$1/3$ & $1.5$ & $3000$ & $16$ & $48$ & met & $50$ & $1.55(31)$\\
$1/3$ & $1.51$ & $4749$ & $16$ & $48$ & met & $100$ & $1.29(20)$\\
$1/3$ & $1.515$ & $15000$ & $16$ & $48$ & met & $50$ & $2.75(36)$\\
$1/3$ & $1.55$ & $4749$ & $16$ & $48$ & met & $100$ & $1.60(27)$\\
$1/3$ & $1.7$ & $6000$ & $16$ & $48$ & met & $50$ & $2.34(43)$\\
$1/4$ & $1.4$ & $16002$ & $16$ & $64$ & met & $50$ & $5.23(88)$\\
$1/4$ & $1.415$ & $9000$ & $16$ & $64$ & met & $100$ & $10.9(31)$\\
$1/4$ & $1.42$ & $9000$ & $16$ & $64$ & met & $100$ & $3.05(52)$\\
$1/4$ & $1.43$ & $22002$ & $16$ & $64$ & met & $50$ & $5.56(84)$\\
$1/4$ & $1.4378$ & $17002$ & $16$ & $64$ & met & $50$ & $6.6(12)$\\
$1/4$ & $1.45$ & $10000$ & $16$ & $64$ & met & $50$ & $6.9(16)$\\
$1/4$ & $1.46$ & $2499$ & $16$ & $64$ & met & $100$ & $2.90(81)$\\
$1/4$ & $1.47$ & $2499$ & $16$ & $64$ & met & $100$ & $2.92(82)$\\
$1/4$ & $1.49$ & $9000$ & $16$ & $64$ & met & $100$ & $5.3(11)$\\
$1/4$ & $1.5$ & $9000$ & $16$ & $64$ & met & $100$ & $5.1(11)$\\
$1/4$ & $1.53$ & $8999$ & $16$ & $64$ & met & $100$ & $9.8(27)$\\
$1/4$ & $1.7$ & $7999$ & $16$ & $64$ & met & $50$ & $17.7(63)$\\
$1/5$ & $1.39$ & $7502$ & $16$ & $80$ & hb & $50$ & $2.41(40)$\\
$1/5$ & $1.4$ & $7502$ & $16$ & $80$ & hb & $50$ & $2.22(36)$\\
$1/5$ & $1.41$ & $7502$ & $16$ & $80$ & hb & $50$ & $3.09(57)$\\
$1/5$ & $1.42$ & $3500$ & $16$ & $80$ & hb & $50$ & $2.39(54)$\\
$1/5$ & $1.435$ & $7501$ & $16$ & $80$ & hb & $50$ & $3.52(68)$\\
$1/5$ & $1.46$ & $7502$ & $16$ & $80$ & hb & $50$ & $3.52(68)$\\
$1/5$ & $1.47$ & $7502$ & $16$ & $80$ & hb & $50$ & $3.71(74)$\\
$1/5$ & $1.48$ & $7502$ & $16$ & $80$ & hb & $50$ & $4.7(10)$\\
$0.18$ & $1.39$ & $176221$ & $16$ & $88$ & hb & $50$ & $1.55(15)$\\
$0.18$ & $1.395$ & $260004$ & $16$ & $88$ & hb & $50$ & $1.61(13)$\\
$0.18$ & $1.4$ & $176256$ & $16$ & $88$ & hb & $50$ & $1.38(12)$\\
$0.18$ & $1.405$ & $260004$ & $16$ & $88$ & hb & $50$ & $1.64(13)$\\
$0.18$ & $1.41$ & $175633$ & $16$ & $88$ & hb & $50$ & $1.80(18)$\\
$0.18$ & $1.45$ & $176286$ & $16$ & $88$ & hb & $50$ & $2.76(34)$\\
$0.18$ & $1.46$ & $176291$ & $16$ & $88$ & hb & $50$ & $3.20(42)$\\
$0.18$ & $1.47$ & $176288$ & $16$ & $88$ & hb & $50$ & $3.12(40)$\\
\hline
\caption{
List of configurations with $L=16$ that were used to determine the continuum limit.
We give the coupling $\beta$, the input anisotropy $\xiinput$, the number of thermalised configurations on which we did measurements, the number of spatial and temporal lattice points, the algorithm used for generation, the number of sweeps that were done between measurements and the integrated autocorrelation time of the spatial-spatial plaquette.
The algorithm is one of heatbath-overrelaxation (hb) and Metropolis (met).}
\label{tab:usedconfigsL16}

\end{longtable}    \begin{longtable}{llllllll}
    \hline
    $\xiinput$ & $\beta$ & \# confs. & $L$ & $T$ & algo & \# sweeps & $\tau_\text{int}$\\
    \hline
$1$ & $1.65$ & $9667$ & $3$ & $16$ & met & $50$ & $0.4818(99)$\\
$1$ & $1.7$ & $9667$ & $3$ & $16$ & met & $50$ & $0.497(10)$\\
$4/5$ & $1.6$ & $7501$ & $3$ & $20$ & met & $100$ & $0.491(20)$\\
$4/5$ & $1.615$ & $7501$ & $3$ & $20$ & met & $100$ & $0.489(23)$\\
$4/5$ & $1.64$ & $7501$ & $3$ & $20$ & met & $100$ & $0.522(36)$\\
$4/5$ & $1.65$ & $9667$ & $3$ & $20$ & met & $50$ & $0.500(20)$\\
$4/5$ & $1.6521$ & $7501$ & $3$ & $20$ & met & $100$ & $0.490(11)$\\
$2/3$ & $1.56$ & $7251$ & $3$ & $24$ & met & $100$ & $0.481(11)$\\
$2/3$ & $1.5614$ & $7251$ & $3$ & $24$ & met & $100$ & $0.502(33)$\\
$2/3$ & $1.565$ & $7251$ & $3$ & $24$ & met & $100$ & $0.496(31)$\\
$2/3$ & $1.58$ & $9167$ & $3$ & $24$ & met & $50$ & $0.501(23)$\\
$2/3$ & $1.595$ & $46250$ & $3$ & $24$ & hb & $20$ & $0.4944(80)$\\
$2/3$ & $1.6$ & $7251$ & $3$ & $24$ & met & $100$ & $0.498(31)$\\
$2/3$ & $1.6075$ & $7251$ & $3$ & $24$ & met & $100$ & $0.499(12)$\\
$1/2$ & $1.49$ & $9667$ & $3$ & $32$ & met & $100$ & $0.496(10)$\\
$1/2$ & $1.515$ & $9667$ & $3$ & $32$ & met & $100$ & $0.493(10)$\\
$1/2$ & $1.5381$ & $9667$ & $3$ & $32$ & met & $100$ & $0.607(44)$\\
$1/2$ & $1.54$ & $8667$ & $3$ & $32$ & met & $50$ & $0.493(21)$\\
$1/2$ & $1.55$ & $8667$ & $3$ & $32$ & met & $50$ & $0.498(11)$\\
$1/2$ & $1.5514$ & $9667$ & $3$ & $32$ & met & $100$ & $0.532(32)$\\
$1/2$ & $1.5531$ & $8667$ & $3$ & $32$ & met & $50$ & $0.634(51)$\\
$1/2$ & $1.555$ & $9667$ & $3$ & $32$ & met & $100$ & $0.491(10)$\\
$2/5$ & $1.46$ & $14501$ & $3$ & $40$ & met & $100$ & $0.502(24)$\\
$2/5$ & $1.4638$ & $14501$ & $3$ & $40$ & met & $100$ & $0.724(51)$\\
$2/5$ & $1.47$ & $14501$ & $3$ & $40$ & met & $100$ & $0.507(22)$\\
$2/5$ & $1.48$ & $14501$ & $3$ & $40$ & met & $100$ & $0.667(45)$\\
$2/5$ & $1.51$ & $7667$ & $3$ & $40$ & met & $50$ & $0.573(43)$\\
$2/5$ & $1.511$ & $7667$ & $3$ & $40$ & met & $50$ & $0.524(34)$\\
$2/5$ & $1.52$ & $14501$ & $3$ & $40$ & met & $100$ & $0.539(28)$\\
$2/5$ & $1.525$ & $14501$ & $3$ & $40$ & met & $100$ & $0.729(51)$\\
$1/3$ & $1.44$ & $37001$ & $3$ & $48$ & met & $100$ & $0.605(24)$\\
$1/3$ & $1.4406$ & $37001$ & $3$ & $48$ & met & $100$ & $1.407(93)$\\
$1/3$ & $1.45$ & $37001$ & $3$ & $48$ & met & $100$ & $1.310(83)$\\
$1/3$ & $1.46$ & $4667$ & $3$ & $48$ & met & $50$ & $0.791(95)$\\
$1/3$ & $1.48$ & $4667$ & $3$ & $48$ & met & $50$ & $0.93(12)$\\
$1/3$ & $1.4814$ & $37001$ & $3$ & $48$ & met & $100$ & $1.326(85)$\\
$1/3$ & $1.5$ & $37001$ & $3$ & $48$ & met & $100$ & $1.292(82)$\\
$1/4$ & $1.42$ & $49334$ & $3$ & $64$ & met & $100$ & $9.6(13)$\\
$1/4$ & $1.43$ & $29336$ & $3$ & $64$ & met & $50$ & $4.53(55)$\\
$1/4$ & $1.4378$ & $2667$ & $3$ & $64$ & met & $50$ & $2.39(61)$\\
$1/4$ & $1.47$ & $49334$ & $3$ & $64$ & met & $100$ & $1.72(11)$\\
$1/4$ & $1.49$ & $49334$ & $3$ & $64$ & met & $100$ & $12.8(19)$\\
$1/5$ & $1.41$ & $47499$ & $3$ & $80$ & hb & $200$ & $0.820(37)$\\
$1/5$ & $1.42$ & $47500$ & $3$ & $80$ & hb & $200$ & $0.868(40)$\\
$1/5$ & $1.46$ & $47499$ & $3$ & $80$ & hb & $200$ & $0.923(44)$\\
$1/5$ & $1.47$ & $47500$ & $3$ & $80$ & hb & $200$ & $0.964(47)$\\
$1/5$ & $1.48$ & $47500$ & $3$ & $80$ & hb & $200$ & $0.986(48)$\\
$0.18$ & $1.39$ & $36770$ & $3$ & $88$ & hb & $500$ & $1.349(87)$\\
$0.18$ & $1.395$ & $36000$ & $3$ & $88$ & hb & $500$ & $1.284(81)$\\
$0.18$ & $1.4$ & $36000$ & $3$ & $88$ & hb & $500$ & $1.420(95)$\\
$0.18$ & $1.405$ & $19742$ & $3$ & $88$ & hb & $500$ & $1.28(11)$\\
$0.18$ & $1.45$ & $36756$ & $3$ & $88$ & hb & $500$ & $1.69(12)$\\
$0.18$ & $1.46$ & $37277$ & $3$ & $88$ & hb & $500$ & $1.99(15)$\\
$0.18$ & $1.47$ & $36000$ & $3$ & $88$ & hb & $500$ & $1.75(13)$\\
\hline
\caption{
List of configurations with $L=3$ that were used to determine the continuum limit.
We give the coupling $\beta$, the input anisotropy $\xiinput$, the number of thermalised configurations on which we did measurements, the number of spatial and temporal lattice points, the algorithm used for generation, the number of sweeps that were done between measurements and the integrated autocorrelation time of the spatial-spatial plaquette.
The algorithm is one of heatbath-overrelaxation (hb) and Metropolis (met).}
\label{tab:usedconfigsL3}
\end{longtable}

\printbibliography

\end{document}